\documentclass[sigconf,nonacm]{acmart}

\AtBeginDocument{%
  }

\setcopyright{none} 
\usepackage{listings}
\usepackage{colortbl}
\usepackage{multirow}
\usepackage{makecell} % in-cell linefeed
\usepackage{pifont}
\usepackage{algorithm}
\usepackage{algorithmic} 
\usepackage{soul} % \hl

\begin{document}

%%
%% The "title" command has an optional parameter,
%% allowing the author to define a "short title" to be used in page headers.
\title{\texorpdfstring{$\rho$Hammer: Reviving RowHammer Attacks on New Architectures via Prefetching}{rhoHammer: Reviving RowHammer Attacks on New Architectures via Prefetching}}
% \subtitle{\normalsize{MICRO 2025 Submission
%     \textbf{\#206} -- Confidential Draft -- Do NOT Distribute!!}}
%%
%% The "author" command and its associated commands are used to define
%% the authors and their affiliations.
%% Of note is the shared affiliation of the first two authors, and the
%% "authornote" and "authornotemark" commands
%% used to denote shared contribution to the research.
%\author{\normalsize{ISCA 2025 Submission
 %   \textbf{\#NaN} -- Confidential Draft -- Do NOT Distribute!!}}

%%
%% By default, the full list of authors will be used in the page
%% headers. Often, this list is too long, and will overlap
%% other information printed in the page headers. This command allows
%% the author to define a more concise list
%% of authors' names for this purpose.

\author{Weijie Chen}
\affiliation{%
  \institution{Huazhong Univ. of Sci. and Tech.}
  \city{Wuhan}
  \country{China}
}
\affiliation{%
  \institution{The Hong Kong Polytechnic Univ.}
  \city{Hong Kong}
  \country{China}
}
\email{weijie_chen@hust.edu.cn}

\author{Shan Tang}
\authornote{Both authors contributed equally to this research.}
\affiliation{%
  \institution{Huazhong Univ. of Sci. and Tech.}
  \city{Wuhan}
  \country{China}
}
\email{m202372164@hust.edu.cn}

\author{Yulin Tang}
\authornotemark[1]
\affiliation{%
  \institution{Huazhong Univ. of Sci. and Tech.}
  \city{Wuhan}
  \country{China}
}
\email{m202372045@hust.edu.cn}

\author{Xiapu Luo}
\affiliation{%
  \institution{The Hong Kong Polytechnic Univ.}
  \city{Hong Kong}
  \country{China}
}
\email{csxluo@comp.polyu.edu.hk}

\author{Yinqian Zhang}
\authornote{Yinqian Zhang is affiliated with Research Institute of Trustworthy Autonomous Systems and the Department of Computer Science and Engineering of SUSTech.}
\affiliation{%
  \institution{Southern Univ. of Sci. and Tech.}
  \city{Shenzhen}
  \country{China}
}
\email{yinqianz@acm.org}

\author{Weizhong Qiang}
\authornote{Weizhong Qiang is the corresponding author and affiliated with Jinyinhu Laboratory, Hubei Key Laboratory of Distributed System Security, and Hubei Engineering Research Center on Big Data Security.}
\affiliation{%
  \institution{Huazhong Univ. of Sci. and Tech.}
  \city{Wuhan}
  \country{China}
}
\email{wzqiang@hust.edu.cn}

%%
%% The abstract is a short summary of the work to be presented in the
%% article.

%%%%%% -- PAPER CONTENT STARTS-- %%%%%%%%
\thanks{This is the author’s version of the work accepted at the 58th IEEE/ACM International Symposium on Microarchitecture (MICRO '25). The version published by ACM is at \url{https://doi.org/10.1145/3725843.3756042}.}
\begin{abstract}

  Rowhammer is a critical vulnerability in dynamic random access memory (DRAM) that continues to pose a significant threat to various systems.
  However, we find that conventional load-based attacks are becoming highly ineffective on the most recent architectures such as Intel Alder and Raptor Lake.
  In this paper, we present $\rho$Hammer, a new Rowhammer framework that systematically overcomes three core challenges impeding attacks on these new architectures.
  First, we design an efficient and generic DRAM address mapping reverse-engineering method that uses selective pairwise measurements and structured deduction, enabling recovery of complex mappings within seconds on the latest memory controllers.
  Second, to break through the activation rate bottleneck of load-based hammering, we introduce a novel prefetch-based hammering paradigm that leverages the asynchronous nature of x86 prefetch instructions and is further enhanced by multi-bank parallelism to maximize throughput.
  Third, recognizing that speculative execution causes more severe disorder issues for prefetching, which cannot be simply mitigated by memory barriers, we develop a counter-speculation hammering technique using control-flow obfuscation and optimized \texttt{NOP}-based pseudo-barriers to maintain prefetch order with minimal overhead.
  Evaluations across four latest Intel architectures demonstrate $\rho$Hammer’s breakthrough effectiveness: it induces up to 200K+ additional bit flips within 2-hour attack pattern fuzzing processes and has a 112× higher flip rate than the load-based hammering baselines on Comet and Rocket Lake.
  Also, we are the first to revive Rowhammer attacks on the latest Raptor Lake architecture, where baselines completely fail, achieving stable flip rates of 2,291/min and fast end-to-end exploitation.

\end{abstract}


%%
%% Keywords. The author(s) should pick words that accurately describe
%% the work being presented. Separate the keywords with commas.

\maketitle

\section{Introduction}

Rowhammer is a well-known circuit-level vulnerability within DRAM that exploits the read disturbance effect to flip the value stored in memory cells (a.k.a. activation-induced bit flips, AIB~\cite{DRAMScope}) by repeatedly opening and closing their nearby DRAM rows for sufficient times.
Since its first-time discovery~\cite{Fliping}, this critical issue has drawn profound attention from both industry and academia for over a decade.
At its core, Rowhammer attack enables the manipulation of memory contents without directly accessing them, which has proven its potential to compromise a wide range of systems and application scenarios, such as privilege escalation and model intelligence degradation~\cite{Gogogadget,Drammer,Yes_One-Bit-Flip_Matters!,Flip_Feng_Shui,SpecHamme,When_Frodo_Flips,Another_Flip_in_the_Wall_of_Rowhammer_Defenses,one_Bit_Flips_One_Cloud_Flops,pthammer,deephammer,seaborn2015exploiting,Rowhammerjs,terminal_brain,SGX-Bomb,HammerScope,hyperhammer,ghostknight,halfspectre}, and resulted in significant security implications on system isolation and trusted computing~\cite{hitchhiker,erebor,osdi_ryoan,osdi_blackbox,taco_trusted_serverless,atc_asterinas}.

Currently, it has been widely acknowledged that the Rowhammer issue would keep exacerbating with the evolution of DDR standards and the ever-increasing density of DRAM cells~\cite{Fliping}.
In response, valiant efforts have been made to address this issue~\cite{ZebRAM,Sok,TWiCe,Leveraging_EM_Side_Channel_Information_to_Detect_Rowhammer_Attacks,Cannot_Touch_This,ANVIL,Graphene}, with target row refresh (TRR) currently being the most widely implemented defense mechanism in off-the-shelf DDR4 devices.
TRR works by periodically refreshing a subset of rows being frequently activated, thereby effectively preventing early Rowhammer strategies that relied on simple patterns like the double-sided hammering (i.e., two aggressor rows sandwiching one victim row)~\cite{Micron,Samsung,Micron_inc}.
Unfortunately, recent research~\cite{trrespass,utrr,Half-Double,blacksmith,zenhammer} has demonstrated that the TRR mitigation within DDR4 DRAMs can be effectively bypassed, particularly through the state-of-the-art non-uniform hammering techniques that confuse the TRR sampler in identifying the real victim rows.

Nonetheless, it seems that Rowhammer attacks have not evolved to be more potent over time as anticipated.
Both this paper and recent literature~\cite{zenhammer,sledgehammer} acknowledge that such attacks are becoming increasingly difficult to realize on newer platforms.
After systematically investigating the four most recent Intel architectures, we identify the following non-trivial challenges and propose a novel prefetch-based Rowhammer framework, $\rho$Hammer, which is significantly more effective than conventional load-based attacks and revives Rowhammer on the latest architectures.% previously considered infeasible.

\textbf{Challenge I: Complex DRAM Address Mappings}

A critical prerequisite for launching effective Rowhammer attacks is the ability to reverse-engineer the correct DRAM address mapping of the target system in order to precisely locate the victim rows.
While numerous techniques~\cite{drama,dramdig,zenhammer,one_Bit_Flips_One_Cloud_Flops} have been proposed, they primarily rely on incomplete heuristics that become impractical as architectural complexity increases.
For instance, in this paper, we identify that the latest Alder/Raptor Lake architectures adopt more intricate mapping schemes with expanded bank bit ranges and bank function combinations, which is likely in response to their supporting DDR5 with extra bank groups~\cite{jedec_ddr5,12gen_manual}.
To overcome these limitations, we present a new layout-agnostic reverse-engineering methodology that leverages selective pairwise measurements and structured deduction to accurately and generically recover mappings across architectures.
Meanwhile, it also achieves high efficiency and scalability by addressing the combinatorial explosion issue of prior brute-force-based approaches.
Evaluation across our hardware setups demonstrates that our method recovers complete mappings within 10 seconds while all available tools completely fail on Alder/Raptor Lake.

\textbf{Challenge II: Insufficient Activation Rate}

One major motivation of ours stems from observing that existing Rowhammer methods, which primarily rely on memory load instructions (e.g., x86 \texttt{MOV}) for DRAM accesses and memory barrier instructions to maintain their order~\cite{blacksmith,zenhammer,trrespass,ECC,sledgehammer,SPOILER,WhistleBlower,Are_We_Susceptible_to_Rowhammer}, are not as effective on newer architectures as originally reported, especially on the most recent Alder/Raptor Lake platforms where they consistently fail.
A deeper analysis of this paper reveals that the activation rate of these methods still requires further improvement to make Rowhammer attacks (more) generic and exploitable.
To address this, we introduce a novel prefetch-based hammering paradigm that leverages the asynchronous and non-blocking characteristics of x86 prefetch instructions to significantly enhance the hammering throughput.
In addition, by combining prefetching with the multi-bank technique, we are able to amplify our attacks even further through bank-level parallelism.

\textbf{Challenge III: Speculative Disorder Hazard of Prefetching}

% 和activation rate衔接？diorder会降低实际的ar
Modern speculative execution mechanisms, particularly out-of-order (OoO) execution and branch prediction, present a significant challenge to Rowhammer attacks by disrupting the intended instruction order and thereby reducing the overall hammering effectiveness.
While both load-based and prefetch-based hammering are affected by this issue, we find that prefetch instructions exhibit even more severe disorder, particularly on the latest Alder/Raptor Lake platforms.
Even worse, unlike explicit memory reads, prefetching performs implicit accesses and cannot be reliably ordered following the traditional wisdom of using barrier instructions.
To overcome the disorder issue, we develop a counter-speculation hammering technique that combines runtime control flow obfuscation with carefully tuned \texttt{NOP}-based pseudo-barriers.
This technique mitigates the reordering effects derived from both OoO execution and branch prediction, restoring more controllable prefetching behavior that unlocks the potential of prefetch-based hammering.

Our evaluation demonstrates a substantial leap in attack effectiveness with $\rho$Hammer compared to the conventional load-based Rowhammer baselines.
During our large-scale 2-hour fuzzing processes, we observe a maximum of over 200K additional bit flips on Comet Lake alone.
Even on newer Alder/Raptor Lake platforms where the baseline fails to produce any meaningful bit flips, $\rho$Hammer consistently induces thousands.
Moreover, repeated attack simulations across diverse physical locations reveal $\rho$Hammer's high stability and practicality: $\rho$Hammer achieves sustained flip rates of 187K/min and 47K/min on Comet/Rocket Lake, representing 112.4x and 47.1x improvements over the baseline.
On Alder/Raptor Lake, it further maintains highly practical rates of 995/min and 2,291/min, while the baseline could reproduce none.

% \textbf{Summary of Contributions.}
The main contributions of this paper are as follows:
\begin{itemize}
    \item 
    We propose a generic and efficient DRAM address mapping reverse-engineering method that uese structured pairwise measurements without any prior knowledge and assumptions about the mapping pattern, achieving full mapping recovery within 10 seconds across all tested recent platforms.
    \item
    We introduce a novel prefetch-based Rowhammer attack paradigm that significantly boosts activation rate by leveraging the asynchronous nature of prefetch instructions, further enhanced by combining it with multi-bank parallelism.
    \item 
    We develop a counter-speculation hammering technique that mitigates the intensified disorder effects of OoO execution and branch prediction, enabling more effective prefetch-based Rowhammer on the latest architectures.
    \item 
    We open-source our implementation on Github at \url{https://github.com/rhohammer/rhohammer} and evaluate our attack on four recent Intel architectures and show that it achieves orders-of-magnitude higher in bit flip counts and flip rates compared to load-based baselines, including Alder and Raptor Lake platforms where baselines consistently fail.
\end{itemize}

\textbf{Responsible Disclosure.} We have reported potential security impact of $\rho$Hammer to related parties, including Intel, MSI, and Samsung. The PSIRT of both Intel and MSI have been actively collaborating with us to investigate and address the issue. The MSI PSIRT promptly released a fix addressing the issue.

\section{Background} \label{sec:bg}

\subsection{Dynamic Random Access Memory} \label{sec:bg_dram}

\begin{figure}[t!]
    \hspace{-2mm}
    \centering
    \includegraphics[width=0.48\textwidth]{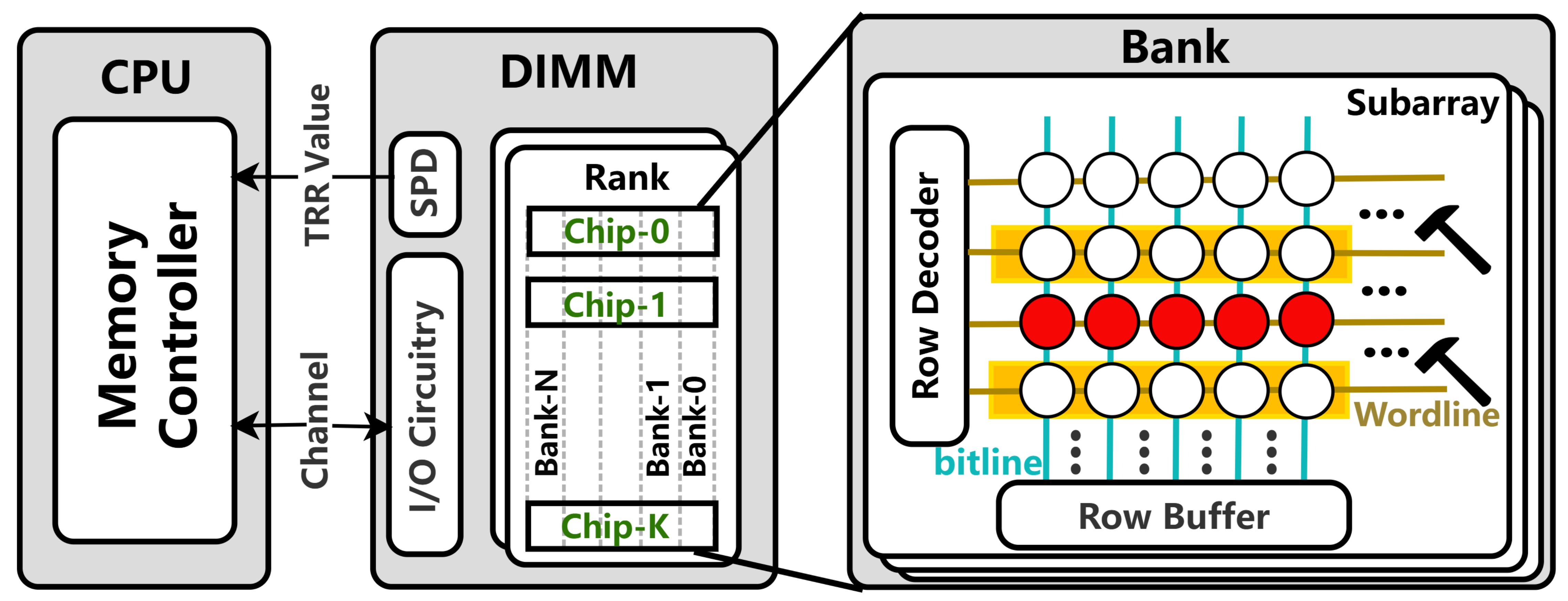}
    \caption{DRAM hierarchical organization.} \label{fig:dram}
    % \Description{Diagram showing the hierarchical organization of DRAM.}
    \vspace{-8mm}
\end{figure}

DRAM is a volatile memory technology widely equipped as the main memory in modern computers, which is usually available on the market in the form of dual inline memory modules (DIMM).
Figure~\ref{fig:dram} shows an overview of the DRAM organization.
A DIMM communicates with the memory controller (MC) through one of the channels and has chips on one or both sides (i.e., single/dual-rank).
Each channel operates independently, while each rank processes commands from the MC in lockstep.
Orthogonally, each chip on a rank is divided into multiple banks (typically 16 banks per rank on DDR4), where each bank contains subarrays that share the same global row buffer and each subarray is a grid of memory cells.
% Such a hierarchical and orthogonal structure is inherently designed to support high parallelism and memory throughput.
Before each DRAM access, one wordline is activated and all cells connected to it are opened, allowing the stored 0/1 value to be loaded to the row buffer, which later serves all subsequent read/write operations.
As capacitors in DRAMs leak charge over time, periodic refreshing is needed to preserve data, where each refresh command refreshes a subset of rows and is issued every $t_{REFI} \approx 7.8\,\mu s$ on average.
% In DDR4, the default refresh window $t_{REFW}$ is 64 ms, which translates to approximately 8192 refreshes (\texttt{REF}) commands.
% More specifically, a \texttt{REF} command refreshes a subset of rows and is issued every $t_{REFI} = 7.9\mu s$ on average.

\textbf{SBDR Side Channel.}
When two subsequent DRAM accesses targeting the same bank but different rows (SBDR), the timing latency will be significantly higher than in other cases, namely same-row (SR) and different-bank (DB), forming the SBDR side channel (a.k.a row conflict side channel).
The fundamental reason lies in the fact that only one row can be loaded into the global row buffer per bank at any given time, which accelerates subsequent accesses hitting the row buffer.
However, to access a second row within the same bank, a precharge (\texttt{PRE}) command must be issued to close the currently open row, introducing an extra latency. % specified in the DDR standard: the minimal interval required between two sequential row activations within the same bank ($t_{RC}$).

\textbf{DRAM Address Mapping.}
In modern computer systems, there exists a well-defined mapping between physical addresses and geographical DRAM addresses, which is essential for the memory controller to efficiently fetch data from the correct DRAM cell.
Specifically, physical addresses are translated into the bank, row, and column addresses, the process of which is governed by the memory controller and thus is CPU-specific.
Accurate knowledge of the DRAM addressing function is essential for effective Rowhammer exploit, especially the row bits and bank functions (typically linear xors of bank bits), but this information is usually kept proprietary in most commercial CPUs~\cite{utrr,trrespass}.
Previous studies have leveraged the SBDR side channel to reverse-engineer such mapping information~\cite{drama,zenhammer,risc-h,one_Bit_Flips_One_Cloud_Flops,dramdig,bitmine,Reliable_Reverse_Engineering_of_Intel_DRAM_Addressing_Using_Performance_Counters}.

% bank-conflict side-channel
\subsection{Rowhammer Attacks} \label{sec:bg_rh}

Rowhammer is a critical DRAM vulnerability that exploits the read disturbance effect, where repeated accesses to a specific memory row (aggressor row) induce bit flips in adjacent rows (victim rows).
This phenomenon occurs because each activation of a row draws a tiny amount of charge from nearby rows through capacitive coupling or electron migration~\cite{DRAMScope}.
Over time, this cumulative charge leakage leads to the state of a memory cell unintentionally flips from 0 to 1 or vice versa.
% With advancements in semiconductor technology, it has been acknowledged that the Rowhammer vulnerability is becoming more severe as the density of DRAM cells continues to increase[].
In response, the industry has widely integrated the TRR mitigation on modern DDR4 devices, which works by detecting potential victim rows during regular refresh operations and proactively refreshing them to prevent bit flips.
While TRR has been effective in mitigating traditional Rowhammer attacks, recent research has successfully bypassed these defenses ~\cite{blacksmith,trrespass,zenhammer}.
Notably, recent non-uniform hammering~\cite{blacksmith,zenhammer} represents the most effective state-of-the-art approach that uses carefully crafted patterns to make the TRR mechanism more difficult to determine the correct victim row.
% These methods involve hammering multiple aggressor rows with varying frequencies, phases, and amplitudes to exploit blind spots in existing mitigations.

\subsection{x86 Prefetch Instructions} \label{sec:bg_pre}

These are a set of specified instructions designed to improve memory access performance by fetching data into the cache before it is actually needed.
In contrast with hardware prefetching widely integrated in modern caches that is completely transparent to users, prefetch instructions (i.e., software prefetching) provide an interface for programmers and compilers to explicitly hint the CPU about future accesses.
One notable performance aspect is that these instructions are asynchronous and do not stall the CPU while the data is being fetched, which allows the CPU to continue executing other tasks concurrently.
The x86 ISA supports several \texttt{PREFETCHh} instructions, namely \texttt{PREFETCHT0} (fetch to all cache levels), \texttt{PREFETCHT1} (fetch to the L2 and LLC), \texttt{PREFETCHT2} (fetch to the LLC), and \texttt{PREFETCHNTA} (non-temporal), in decreasing order of data temporal locality~\cite{intel_arch_opt}.
We exclude the prefetch write instruction \texttt{PREFETCHw} in this paper as it further changes the cache coherence state that may cause extra overhead than \texttt{PREFETCHh} instructions~\cite{adversarial_prefetch}.
% However, the intricate designs of prefetch instructions is subject to different vendors and micro-architectures.
% For example, Intel's \texttt{prefetchnta} instrcution moves data into the LLC if inclusive, and further into the L1-D [].

\section{Reverse-Engineering Complex DRAM Address Mappings} \label{sec:re}

\subsection{Settings}

\begin{table}[t!]
    \setlength{\extrarowheight}{2.5pt} 
    \setlength{\tabcolsep}{11pt} 
    \centering
    \small
    \begin{tabular}{ccc}
        \bottomrule
        \textbf{Arch. Name} & \multicolumn{1}{c}{\textbf{CPU (Intel Core)}} & \multicolumn{1}{c}{\textbf{Max Mem Freq.}} \\
        \hline
        \cellcolor[gray]{0.9}Comet Lake & \cellcolor[gray]{0.9}i7-10700K & \cellcolor[gray]{0.9}2933 \\
        Rocket Lake & i7-11700 & 2933 \\
        \cellcolor[gray]{0.9}Alder Lake & \cellcolor[gray]{0.9}i9-12900 & \cellcolor[gray]{0.9}3200 \\
        Raptor Lake & i7-14700K & 3200 \\
        \bottomrule
    \end{tabular}
    \caption{Desktop machine setups.}
    \label{tab:cpu_setup}
    \vspace{-7mm}
\end{table}

\begin{normalsize}
\begin{table}[t!]
    \setlength{\extrarowheight}{2.5pt}
    \setlength{\tabcolsep}{10pt}
    \centering
    \small
    \begin{tabular}{ccccc}
        \bottomrule
        \textbf{ID} & \makecell{\textbf{Production}\\\textbf{Date}} & \makecell{\textbf{Freq.}\\(MHz)} & \makecell{\textbf{Size}\\(GiB)} & \makecell{\textbf{Geometry}\\(RK, BK, R)} \\
        \hline
        \cellcolor[gray]{0.9}$\mathcal{S}_1$ & \cellcolor[gray]{0.9}W35-2023 & \cellcolor[gray]{0.9}3200 & \cellcolor[gray]{0.9}16 & \cellcolor[gray]{0.9}(2, 16, $2^{16}$) \\
        $\mathcal{S}_2$ & W33-2021 & 3200 & 8 & (1, 16, $2^{16}$) \\
        \cellcolor[gray]{0.9}$\mathcal{S}_3$ & \cellcolor[gray]{0.9}W30-2020 & \cellcolor[gray]{0.9}2933 & \cellcolor[gray]{0.9}16 & \cellcolor[gray]{0.9}(2, 16, $2^{16}$) \\
        $\mathcal{S}_4$ & W49-2018 & 2666 & 16 & (2, 16, $2^{16}$) \\
        \cellcolor[gray]{0.9}$\mathcal{S}_5$ & \cellcolor[gray]{0.9}W22-2017 & \cellcolor[gray]{0.9}2400 & \cellcolor[gray]{0.9}16 & \cellcolor[gray]{0.9}(2, 16, $2^{16}$) \\
        $\mathcal{H}_1$ & W13-2020 & 2666 & 16 & (2, 16, $2^{16}$) \\
        \cellcolor[gray]{0.9}$\mathcal{M}_1$ & \cellcolor[gray]{0.9}W01-2024 & \cellcolor[gray]{0.9}3200 & \cellcolor[gray]{0.9}32 & \cellcolor[gray]{0.9}(2, 16, $2^{17}$) \\
        \bottomrule
    \end{tabular}
    \caption{DDR4 UDIMMs used in this paper. The DRAM vendors are denoted as $\mathcal{S}$/$\mathcal{H}$/$\mathcal{M}$. RK, BK, and R represent the number of ranks, banks, and rows.}
    \label{tab:dimms}
    \vspace{-9mm}
\end{table}
\end{normalsize}

As listed in Table~\ref{tab:cpu_setup}, this paper covers consecutive generations of desktop Intel Core processors that represent the most recent commercial-off-the-shelf architectures, spanning across Comet Lake (10th-gen), Rocket Lake (11th-gen), Alder Lake (12th-gen), and Raptor Lake (14th-gen).
The 13th-gen is skipped as it also primarily uses the Raptor Lake architecture.
% Given that the 10$^{th}$/11$^{th}$-gen and 12/14$^{th}$-gen processors are compatible on the same motherboard but not interchangeable with each other, we setup our experimental machines using two desktop systems to minimize irrelevant effects.
% The 12/14$^{th}$-gen processors and their associated machine components are all purchased brand new from the market.
All machines use a Linux kernel version of 6.8.0-generic with BIOS settings such as device timings, memory frequencies, and hardware prefetchers at their default state.
Although we do not discuss hardware prefetching in this paper as it is much less controllable, it is important to note that software prefetch instructions are considered as hints to such speculative memory fetching behavior of the processors.
For DRAM chips, as listed in Table~\ref{tab:dimms}, we use all brand-new DIMMs recently bought from the market covering the three major DRAM manufacturers.
% 引用寿命影响脆弱度的文章？
In order to better characterize each DIMM separately, we use single-channel configurations throughout this paper.

\textbf{System Requirements.}
Following prior assumptions~\cite{Gogogadget,Drammer,Flip_Feng_Shui,Another_Flip_in_the_Wall_of_Rowhammer_Defenses,seaborn2015exploiting,rambleed,Throwhammer,Dedup_Est_Machina,Grand_unit}, our reverse-engineering process is offline that requires root privileges. %, but no hardware concact is needed
While the x86 cache flush instruction for touching the DRAM is all-user available, we need the root privilege for virtual-to-physical translation information via the \texttt{/proc/pid/pagemap} Linux interface.
Also, we utilize the high-resolution timer (i.e., x86 \texttt{RDTSCP}) for all timing measurements throughout this paper.
Note that while the bank functions can be empirically classified into channel, rank, bank group, and intra-group bank functions, we will not differentiate them in the remainder of this paper, as they basically serve the same role in Rowhammer attacks that change the geographical bank location.
Furthermore, as Rowhammer only relies on row-granularity operations, we hereinafter exclude the discussion of column bits (i.e., the RowPress~\cite{RowPress} effect is out of scope).

\begin{figure}[t!]
    \hspace{-2mm}
    \centering
    \includegraphics[width=0.48\textwidth]{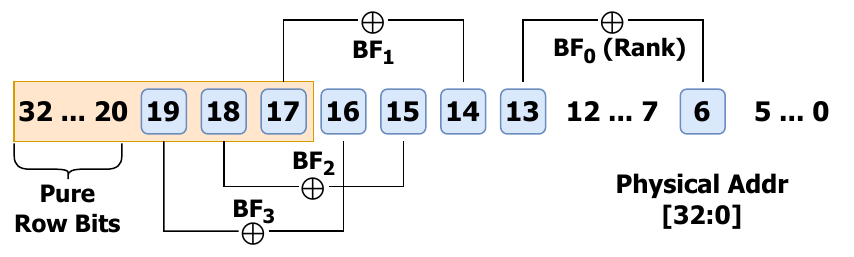}
    \caption{A traditional DRAM address mapping example.} \label{fig:re_map}
    \vspace{-7mm}
\end{figure}

\subsection{Preliminaries} \label{sec:re_prelim}

Figure~\ref{fig:re_map} illustrates a traditional and typical mapping example on the Comet Lake machine, in which the entire physical address space contains row bits (orange), bank bits (blue), and column bits.
In fact, bank bits may share a part with row and column bits, where we refer to \textit{pure row bits} as those that contribute to row addressing alone.
Such patterns are designed to maximize the intra-row and inter-bank utilization, and thus benefit the overall memory throughput.
However, existing reverse-engineering methods rely heavily on heuristics based on established memory mappings, which assume simpler and more predictable patterns of row and bank bits.
Generally, we identify two fundamental sources of limitation in prior works when extending them towards more recent architectures.

\textbf{Complicated Bank Addressing.}
Despite different implementations, the majority of existing approaches~\cite{drama,dramdig,zenhammer} essentially utilize a brute-force-based method that basically assigns each address to a bank (i.e., coloring) and exhausts all possible bank functions, as initially proposed by DRAMA~\cite{drama}.
These methods were effective on older architectures, where the bank function was relatively simple and could be quickly determined.
However, as more complex modern DRAM mappings incorporate more per-function and total bank bits, brute-forcing the bank functions becomes increasingly non-scalable.
Empirically, the total number of candidate bank bits can increase from 19 to 32 and the maximum number of per-function bank bits can expand from 2 to 7, which would result in an 8,192x increase in the total search space and a staggering 23,800x increase in function search overhead.

\begin{algorithm}[t!]
    \caption{Reverse-Engineering Process of $\rho$Hammer}
    \label{alg:re}
    \begin{flushleft}
    \textbf{Input:} $M$: allocated large physical address pool \\
    \textbf{Output:} $bank\_funcs$: recovered set of bank functions; \\ \hspace{30pt} $B_{row}$: recovered set of row bits
    \end{flushleft}
    
    \begin{algorithmic}[1]
        \STATE $thres = $ get\_thres\_prob\_dist$(M)$
        \STATE $bank\_funcs,B_{row} = \{\}$
        \STATE $B_{pure\_row},B_{non\_pure\_row}$ = exclude\_pure\_row\_bits$(M)$
        \FOR{$b_x,b_y \in B_{non\_pure\_row}$}
            \IF{avg$(T_{SBDR}(M,\{b_x,b_y\})) > thres$}
                \STATE $bank\_funcs \gets (b_x,b_y)$ // \textbf{Duet}
            \ENDIF
        \ENDFOR
        \STATE $B_{row} =\ $collect\_higher$(bank\_funcs) \cup B_{pure\_row}$
        \STATE $B_{non\_row} = \{\}$
        \STATE select $\forall\ (b_{BF},b_{BF}') \in bank\_funcs$
        \FOR{$b_x \in B_{non\_pure\_row} \land b_x \notin B_{row}$}
            \IF{avg$(T_{SBDR}(M,\{b_{BF},b_{BF}',b_x\})) < thres$}
                \STATE $B_{non\_row} \gets b_x$ // \textbf{Trios}
            \ENDIF
        \ENDFOR
        \FOR{$b_x,b_y \in B_{non\_row}$}
            \IF{avg$(T_{SBDR}(M,\{b_{BF},b_{BF}',b_x,b_y\})) > thres$}
                \STATE $bank\_funcs \gets (b_x,b_y)$ // \textbf{Quartet}
            \ENDIF
        \ENDFOR
        \STATE merge$(bank\_funcs)$
        \RETURN $bank\_funcs, row\_bits$
    \end{algorithmic}
\end{algorithm}

\textbf{Expired Layout Assumptions.}
In addition, some studies also attempt to enhance the reverse-engineering process by exploiting existing domain knowledge on bit positions, such as excluding pure row bits (see Figure~\ref{fig:re_map}) in the first place to narrow the brute-force search space~\cite{dramdig}, as well as implicitly presume that all bank functions must include row bits~\cite{one_Bit_Flips_One_Cloud_Flops}.
Nevertheless, as we have observed from newer mappings (to be demonstrated), the former strategy is rendered ineffective as pure row bits no longer exist and the latter would overlook low-order functions that cannot be directly derived through the SBDR timings.

To overcome the foregoing limitations, we devise a more generic and efficient reverse-engineering method across architectures.
Different from previous efforts, our technique achieves fast DRAM address mapping recovery with \emph{no} a-priori assumptions such as 1) the total number of bank bits, 2) the size of individual bank functions, or 3) the overlapping relationship between bank and row bits.
Instead, it treats the entire physical address vector as an unknown search space and incrementally infers the mapping in a structured deductive manner, which also effectively reduces the complexity from the exponential brute-forcing to polynomial time.

\subsection{Design} \label{sec:re_design}

Algorithm~\ref{alg:re} shows an overview of our reverse-engineering process.
Throughout the entire process (line 4/12/17), we use a pairwise timing primitive defined as $T_{SBDR}(M, B_{diff})$: the DRAM access timing of an arbitrary address pair selected from the allocated memory pool $M$, where the pair addresses differ only in the bits specified by the set $B_{diff}=\{b_i\}$, while other bits remain identical.
Each primitive is derived from the average of 16 random address pairs, each accessed 50 times, and then compared with a threshold to determine whether it indicates a slower SBDR timing or not.

Without loss of generality, we break down the mapping recovery into three sub-objectives:
1) \textbf{O1}: identify bank functions that include row bits (i.e., row-inclusive functions);
2) \textbf{O2}: determine the range of row bits;
3) \textbf{O3}: identify bank functions that exclude row bits (i.e., non-row functions).
Then we elaborate on each step of the algorithm and how it addresses the above objectives.

\begin{figure}[h]
    \vspace{-3mm}
    \hspace{-2mm}
    \centering
    \includegraphics[width=0.48\textwidth]{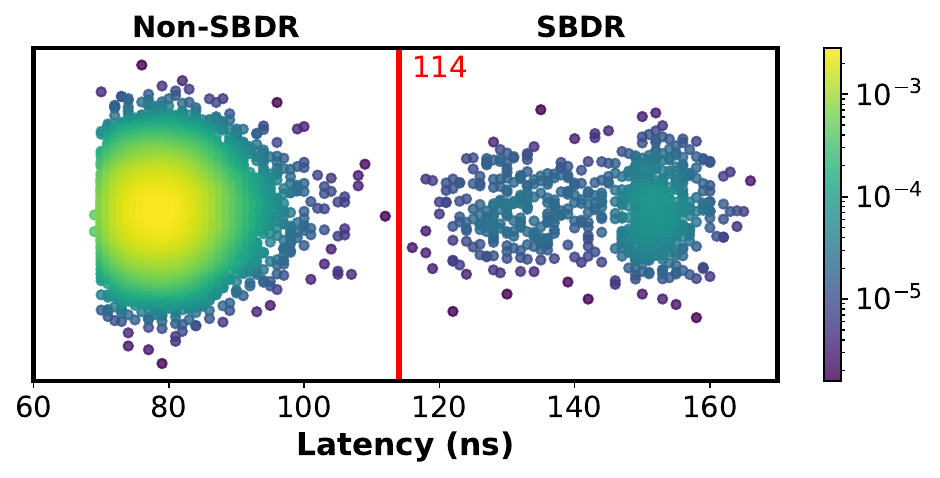}
    \caption{Top-down density distribution map of access latencies with an example threshold; the color scale indicates the proportion of address pairs.} \label{fig:SBDR}
    \vspace{-3mm}
\end{figure}

\textbf{Step 0: Finding the SBDR Threshold.}
The process begins by allocating 4KB pages that occupy more than half (we set this proportion to 70\%) of the system memory to ensure that we can cover the entire physical address space without missing any potential bank bit.
Then, we utilize Linux's pagemap interface to retrieve the page frame numbers of all allocated addresses and maintain their virtual-to-physical mappings, which allows us to quickly select virtual address pairs for $T_{SBDR}$ measurements.
Essential to the reliability of the entire process, a reasonable threshold is needed to distinguish slower SBDR timings from normal ones (line 1).
Therefore, we employ a probability distribution-based method that randomly selects address pairs from (a subset of) our previously allocated memory region.
Figure~\ref{fig:SBDR} shows an example of the distribution of two assembly areas representing SBDR and non-SBDR pairs, whose ratio is approximately $\frac{1}{\#Banks-1}$.
Notably, this ratio is the most accurate if the memory region is aligned to a power-of-two and is large enough to cover at least one bit (i.e., the lowest one) of each bank function.
In this way, its total amount of addresses will be uniformly distributed across all banks, because each bank function will evenly cut the region into 0/1-value halves.

\begin{figure}[t!]
    \hspace{-2mm}
    \centering
    \includegraphics[width=0.48\textwidth]{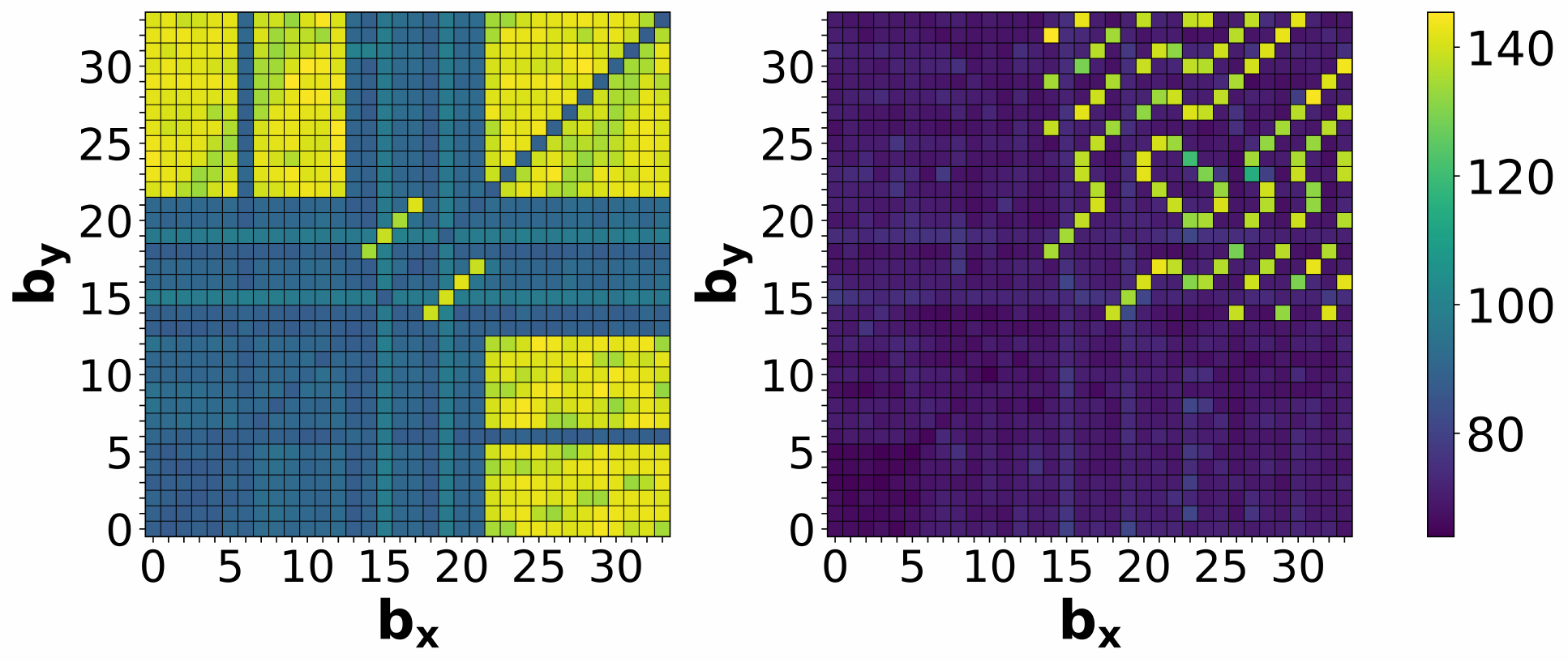}
    \caption{Heatmap of $T_{SBDR}(M, \{b_x,b_y\})$ in \textit{ns} on Comet Lake (left) and Raptor Lake (right), representing traditional and recent mappings, respectively.} \label{fig:re_heatmap}
    \vspace{-6mm}
\end{figure}

\textbf{Step 1: Duet.}
To identify all \emph{row bits and their related bank functions}, we set $B = \{b_x,b_y\}$ and exhaust all combinations of $T_{SBDR}$.
Figure~\ref{fig:re_heatmap} presents the heatmap results illustrating the significant mapping difference across CPU generations, as discussed in Section~\ref{sec:re_prelim}.
%The highlighted blocks indicate that accessing the address pair with $b_x$ and $b_y$ both different incurs the slower SBDR timing.
The highlighted blocks indicate slower SBDR timings when accessing pairs where both $b_x$ and $b_y$ differ.
In the left heatmap, it is noticeable that large chunks of highlighted areas are prevalent, which is due to the presence of high-order pure row bits.
When $b_x$ and $b_y$ are both pure row bits (top right area), or one of them is a pure row bit while another is a pure column bit (top left and bottom right areas), there will be SBDR timings as the row is changed but the bank is not.
Conversely, the missing of such large SBDR chunks in the right heatmap indicates the absence of pure row bits, as their prevalence is now masked by the faster DB timings.

In either case, we first leverage such difference to identify all pure row bits, if any (line 3).
Then, for the remaining bits, we scan over $(b_x,b_y)$ pairs for those with SBDR timings as bank functions, which are visualized as scattered blocks across the right heatmap and in the middle of the left heatmap (line 4-8).
This is because if and only if $b_x$ and $b_y$ reside in the same bank function (i.e., when even number of bits are different, their xor-ed value remains unchanged) and at least one of them is a row bit, we can observe an SBDR timing.
In other words, the combination of all such SBDR pairs forms the complete set of row-inclusive functions (\textbf{O1} resolved).
Meanwhile, after adding up all higher bits in these SBDR pairs, we can reveal all row bits related to bank functions.
Therefore, combining the previously identified pure row bits, this step also recovers the range of row bits (line 9, \textbf{O2} resolved). 

\textbf{Step 2: Trios.}
Evidently, from the heatmap of the previous step, we could not observe any information from the low-order part (bottom left blue areas) of the heatmap, because bank functions there cannot change the row to form an SBDR state, such as $(6,13)$ in Figure~\ref{fig:re_map}.
Therefore, to identify all \emph{remaining non-row functions}, we continue to further extend the measurements with larger $B_{diff}.size$.
Specifically, in this step, we derive all non-row function bits by setting $B_{diff} = \{b_{BF},b_{BF}',b_x\}$, where $(b_{BF},b_{BF}')$ is an arbitrarily chosen bank function pair obtained from the previous step (line 11).
In other words, we have "borrowed" an initial SBDR state from a row-inclusive function.
Then, when we traverse all remaining non-row $b_x$ and measure $T_{SBDR}$, there will be two cases: 1) if $b_x$ is a bank bit, there will be fast DB timings; 2) if $b_x$ is a non-bank bit, the timing would be slow as it still reflects the initial SBDR state.
Consequently, we can distinguish whether $b_x$ is a non-row bank bit and insert it to $B_{non\_row}$ (line 12-16).
% Therefore, we can distinguish whether $b_x$ is a bank bit even if it is a non-row bit because $b_{BF}$ and $b_{BF}'$ have already changed the row indices.

\textbf{Step 3: Quartet.}
Finally, we need to determine all non-row functions using $B_{diff}=\{b_{BF},b_{BF}',b_x, b_y\}$ and traverse all possible combinations of $b_x$ and $b_y$ from $B_{non\_row}$.
Here, we will observe SBDR timings if and only if $b_x$ and $b_y$ reside in the same bank function, because for $b_x, b_y \in BF' \land BF \neq BF'$, there will be even number of different bits in both bank functions (line 17-21, \textbf{O3} resolved).
Lastly, we merge all pairs in $bank\_funcs$ (e.g., $(12,19),(8,12)\xrightarrow{}(8,12,19)$) to construct all bank functions (line 22).
Now, the entire reverse-engineering process completes as we have achieved all sub-objectives, and the evaluation results are later presented in Section~\ref{sec:eval_re}.

\label{F-Q1-1}
With $B_{diff}.size = 4$, we find the algorithm already capable of recovering all existing DDR4 mappings with full accuracy, but further expanding the size and combinations of $B_{diff}$ can provide extra cross-validation.
Due to its layout-agnostic design with polynomial complexity, we believe the method would still preserve its compatibility and scalability towards future mappings even if they further complicate or enlarge the bank search space (e.g., supporting larger physical memory size or using more bank bits).

\section{Hammering Down New Architectures via Counter-Speculation Prefetching} \label{sec:rho}

\subsection{Overview} \label{sec:rho_overview}

After retrieving correct DRAM address mappings, we design our novel prefetch-based Rowhammer paradigm extending from the state-of-the-art non-uniform hammering frameworks~\cite{blacksmith,zenhammer} to advance such attacks on new architectures, which generally addresses two pivotal challenges.
The first one lies in the fundamental limitations of conventional load-based hammering regarding the activation rate.
While it has been proven partially successful on earlier platforms, we notice a considerable gap to make Rowhammer attacks effective on modern architectures.
In response to this, we introduce a novel prefetch-based hammering paradigm that leverages prefetch instructions to substantially enhance hammering throughput.
\label{reb:F-Q3-1}
While prior work~\cite{zenhammer} evaluated the activation rate of using \texttt{PREFETCHNTA}, it attributed the observed improvements solely to increased cache hits and suggested not to use such instructions for hammering.
Alongside, we uncover that the recent multi-bank hammering technique~\cite{sledgehammer} can further boost the efficiency of prefetch-based attacks, as we have seen the same TRR-bypassing patterns yield significantly more bit flips when combined with prefetching and multi-bank techniques.

More importantly, the other challenge pertains to the impact of modern speculative execution mechanisms, which disrupt the intended access order of hammering patterns.
% This disturbance severely undermines the effectiveness of TRR-bypassing patterns and diminishes attack reliability.
Notably, we observe that prefetches exhibit even more pronounced disorder issues compared to loads across all platforms.
Furthermore, since prefetches are not explicit load operations, traditional wisdom such as memory barriers cannot be directly applied to resolve the prefetching disorder issue.
Regarding this, we also note an additional benefit of the multi-bank technique that partially alleviates the disorder effect as its bank-interleaving nature adds pipeline latency to per-bank hammering.
However, as more recent architectures feature intensified speculative behavior, we further introduce a counter-speculation technique to fully unlock the potential of prefetch-based hammering and ensure effectiveness on newer platforms.
By integrating runtime control flow obfuscation and intricate \texttt{NOP}-based pseudo-barriers, we strike the optimal balance between introducing extra overhead and mitigating the negative speculation impacts, thus maximizing the overall effectiveness of our approach.

\begin{figure}[t!]
    % \vspace{-5mm}
    \hspace{-2mm}
    \centering
    \includegraphics[width=0.48\textwidth]{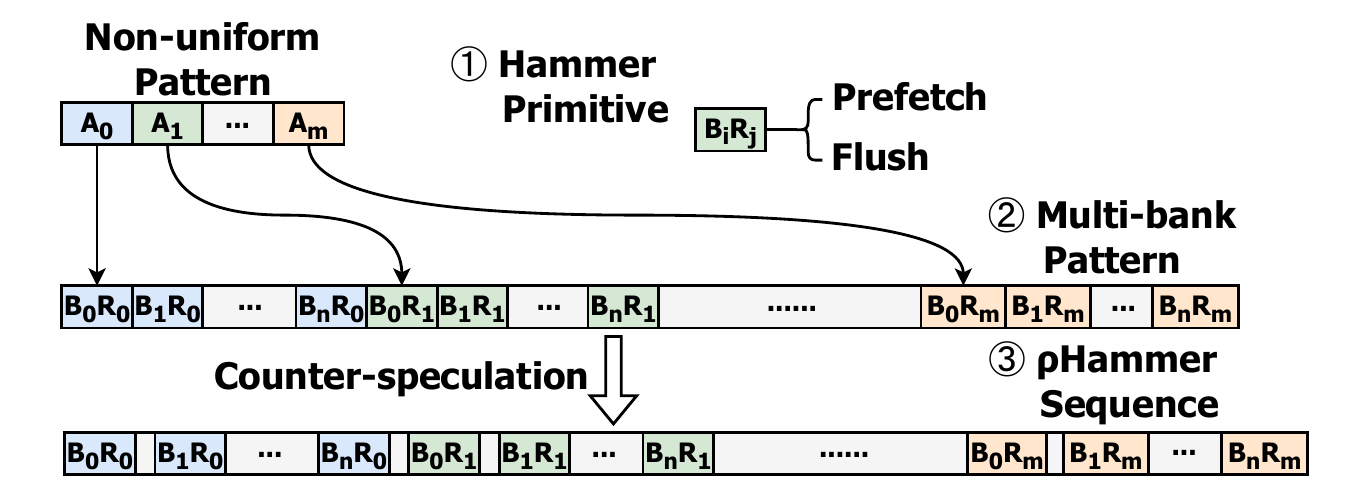}
    \caption{Overview of $\rho$Hammer's workflow. B/R refers to bank/row.} \label{fig:pattern}
    \vspace{-8mm}
\end{figure}

Figure~\ref{fig:pattern} illustrates our general workflow.
Recent Rowhammer attacks against TRR-protected DIMMs typically require executing effective non-uniform hammering patterns - a carefully crafted, ordered sequence of aggressor rows ($A_0 \ldots A_m$).
\label{A-Q2-1}
As noticed by prior work~\cite{blacksmith,posthammer}, the TRR sampler may only observe a subset of activations within a $t_{REFI}$ or not sample from every $t_{REFI}$.
Therefore, such patterns are constructed by assigning varied access frequencies, repetitions, and offsets to a large set of aggressors, attempting to find patterns that effectively confuse the sampler by hiding the true aggressors periodically.
Building upon this concept, $\rho$Hammer incorporates three key techniques, which we will elaborate in the following subsections.

\textbf{Section~\ref{sec:rho_prefetch}:} $\rho$Hammer adopts x86 prefetch instructions as the instrumental hammering primitive (\ding{172}), exploiting their asynchronous nature to increase memory access throughput.

\textbf{Section~\ref{sec:rho_bank}:} $\rho$Hammer distributes each aggressor row across multiple DRAM banks (\ding{173}), allowing the non-uniform attack process to further take advantage of the increased parallelism and alleviated disorder effects.

\textbf{Section~\ref{sec:rho_counter}:} $\rho$Hammer addresses the more pronounced disorder issues of prefetching on the latest architectures by using a counter-speculation hammering technique that intricately inserts runtime latencies to mitigate the branch prediction and OoO effects while minimizing extra overhead to fully realize the potential of prefetch-based hammering throughput (\ding{174}).

\label{reb:C-Q2-1}
\textbf{Threat Model.}
% A complete end-to-end Rowhammer exploit generally consists of three phases: 1) memory templating, 2) memory massaging, and 3) final exploitation~\cite{rubicon}.
% This paper mainly focuses on the first step, where the attacker needs to identify the number and location of bit flips on the system (i.e., perform wide-range sweeping) as they are usually reproducible.
% This step is the most crucial as it directly determine both the attack surface and the vulnerability exposure of real-world DRAM modules, which our newly proposed hammering technique aims to explore.
% Moreover, the difficulty of finding exploitable bit flips in this phase heavily influence the overall runtime and feasibility of practical attacks.
% For convenient and comprehensive characterization on various setups, we do not enforce strict end-to-end attack constraints (e.g., native without \texttt{sudo}), but the proposed hammering strategy is fully compatible with online templating and can be seamlessly integrated into existing massaging and exploitation techniques~\cite{SpecHamme,rubicon,sledgehammer}.
Similar to reverse-engineering, hammering experiments presented in this section do not enforce strict end-to-end attack constraints, such as native without \texttt{sudo} and superpages to facilitate convenient and comprehensive characterization on various setups.
However, as we will show in Section~\ref{sec:eval_sweep}, our proposed attack strategy is able to greatly amplify the chances of templating useful bit flips and can be seamlessly integrated with end-to-end massaging and exploitation~\cite{SpecHamme,rubicon,sledgehammer} techniques.

\textbf{Fuzzing \& Sweeping.}
Following original studies~\cite{blacksmith, zenhammer}, we briefly explain two key operations indicative of non-uniform hammering effectiveness, which will be used for the systematic characterizations in the subsequent sections.
The \textit{fuzzing} process leverages a fuzzer to generate pseudo-random and unique non-uniform patterns, from which \textit{effective patterns} are identified based on the presence of any bit flips during a few execution trials at different physical locations, with the \textit{best pattern} being the one that induces the maximum number of flips.
As each pattern only encodes the relative offsets among aggressor rows, the \textit{sweeping} operation applies an effective pattern repeatedly at a much larger range of different physical locations, which simulates the templating process of real exploits.
While fuzzing primarily reflects the overall effectiveness of a given attack strategy by revealing its difficulty in finding TRR-bypassing patterns, sweeping focuses on the reliability and practical viability of deploying such attacks under general conditions.

\subsection{Potential \& Disorder Issue of Prefetching} \label{sec:rho_prefetch}

As Rowhammer attacks are basically activating and precharging DRAM rows as fast as possible, the activation rate is one of the most critical vectors to measure the hammering effectiveness, as it indicates the highest possible amount of hammer attempts within a DRAM refresh interval.
% the theoretical maximum of which is subject to $t_{RC}$.
Nevertheless, a critical aspect overlooked by past hammering approaches, which only use memory load instructions (e.g., x86 \texttt{MOV}), is that Rowhammer attacks focus solely on the process of retrieving data from DRAM, rather than the final destination.
In other words, the overall activation rate ought to benefit from a shorter lifecycle of the accessed DRAM data.
Prefetch instructions align perfectly with this objective, as their asynchronous nature enables faster execution than normal loads and they only place data in specific caches instead of farther CPU registers, resulting in a shorter data path.
% Therefore, a key motivation of this paper derives from leveraging prefetch instructions to boost the hammering throughput.

\begin{figure}[h]
\centering
\lstset{
  basicstyle=\ttfamily\small,
  frame=single,
  captionpos=b
}
\begin{lstlisting}[caption={A brief overview of the original C++ hammering primitive without any barriers. The hammering primitive is either load-based or prefetch-based.}, label={lst:cpp}]
  // synchronize with refresh command...
  // flush all addresses in aggr_row_addrs[]
  // start hammer pattern:
  for(int idx = 0; idx < num_of_act; idx++){
      hammer(aggr_row_addrs[idx]);
      clflushopt(aggr_row_addrs[idx]);
  }
\end{lstlisting}
\vspace{-5mm}
\end{figure}

\begin{figure}[t!]
    \hspace{-3mm}
    \centering
    \includegraphics[width=0.48\textwidth]{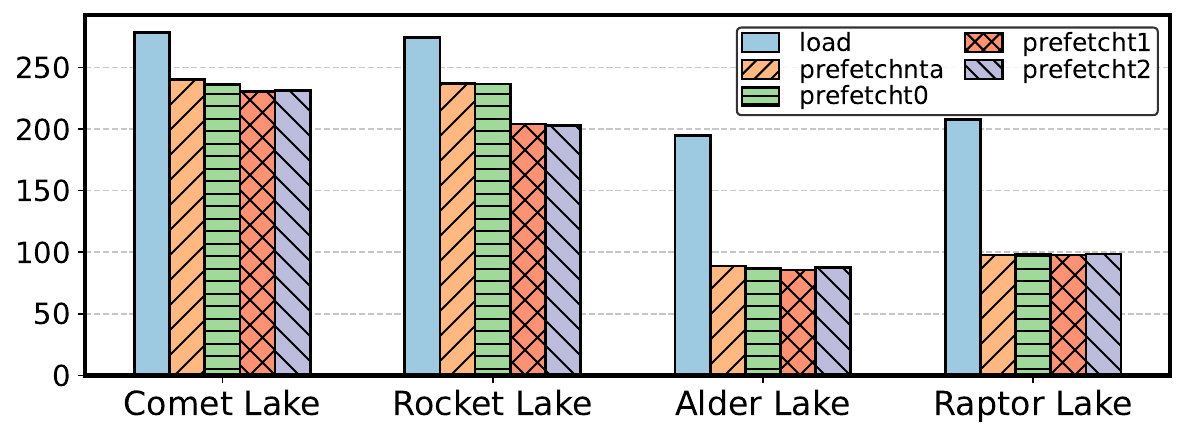}
    \caption{Average attack completion time in \textit{ms} using load or prefetch as the hammering primitive.} \label{fig:time}
    \vspace{-7mm}
\end{figure}

In this section, we focus on exploring the performance potential of prefetch-based hammering compared to conventional load-based hammering, while also revealing the hidden disorder hazard behind the hammering speed boost enabled by prefetch instructions.
We begin by comparing the attack efficiency using a typical hammering primitive shown in Listing~\ref{lst:cpp} with different hammer instructions.
For each platform, we randomly select 80 generated patterns and iteratively execute them until each pattern hits 5 million accesses.
Figure~\ref{fig:time} illustrates the average attack time per pattern using loads and four types of prefetch instructions across the four architectures.
Despite that different prefetch instructions place data in different numbers and levels of caches, their actual attack time shows little variation, yet is still substantially lower than the load-based counterparts.
% We uniformly refer to all variants as prefetch-based in the remainder of this paper.
In the remainder of this paper, $\rho$Hammer empirically uses \texttt{PREFETCHT2} or \texttt{PREFETCHNTA} for prefetch-based hammering that yields marginally better attack performance as they only place data in one level of cache and feature the least temporal locality hint that would minimize unnecessary cache pollution.

However, aligning with prior work~\cite{zenhammer}, we find that the observed advantage in attack time also implicitly includes the increased cache hits caused by the CPU's speculative behavior.
Speculative execution is a critical feature of modern high-performance processors, designed to improve execution speed by issuing instructions in advance rather than strictly following the program order.
Two dominant techniques of speculative execution are out-of-order (OoO) execution and branch prediction.
Although they aim to resolve pipeline stalling issues regarding different aspects, they both rearrange the hammering instruction order that severely affect the attack effectiveness.
Initially, we gain insights into such an impact by looking into the disorder difference between the C++ primitive shown in Listing~\ref{lst:cpp} and its loop-unrolled assembly variant generated via AsmJit~\cite{asmjit}, without any barrier instructions.
Although functionally equivalent, we notice a sharp decline of bit flips after switching from the C++ to the even faster AsmJit primitive on Comet/Rocket Lake machines.
To further investigate, we count their average cache miss rate using the x86 hardware performance counters (HPC)\footnote{Cache misses are measured via the Linux Perf library's \textit{L1-dcache-load-misses} event during the hammer loop.} and the total execution time, where we also observe a dramatic decrease of both when using the AsmJit variant, as shown in the leftmost bars of Figure~\ref{fig:cachemiss}.
By analyzing their dumped binaries, we identify the root cause: just-in-time compiling uses immediate addresses for hammer and flush instructions at each unrolled loop, which allows the CPU to execute aggressively without order.
In contrast, the original C++ primitive features an \texttt{idx} counter serving as a load dependency for indirect addressing at lines 4 and 5 in Listing~\ref{lst:cpp}, which creates a pipeline latency that partially neutralizes the speculative effect.

\begin{figure}[th]
    \centering
    \includegraphics[width=0.48\textwidth]{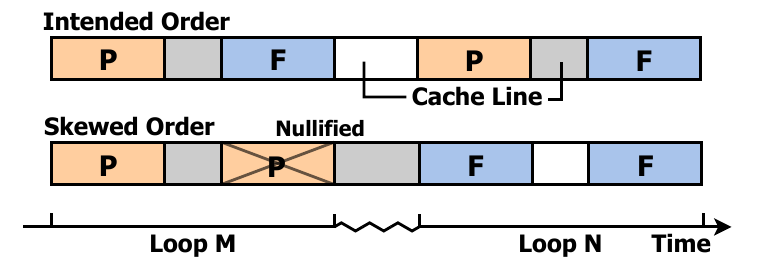}
    \caption{Visualization of prefetch (P) and flush (F) instruction interaction on the same cache line across close hammer loops, where white/gray blocks represent the line absent/present in the cache.} \label{fig:disorder}
    \vspace{-6mm}
\end{figure}

More importantly, for both types of primitives, the cache miss rate of prefetch-based is much lower than that of load-based hammering.
A deeper reason for such a phenomenon lies in the asynchronous nature of prefetch instructions that retire as soon as after finishing the address translation and forwarding a request to the L1-D's line fill buffer (LFB), then they do not stall the subsequent execution while waiting for the pending cache line to be fetched from DRAM and stored in the cache level specified by the hint~\cite{How_to_Be_Fast_and_Not_Furious}.
Also, according to the Intel Software Developer's Manual~\cite{intel_soft_dev}, flush instructions and prefetching behavior are not ordered with each other.
Consequently, as visualized in Figure~\ref{fig:disorder} when a prefetch instruction is issued close enough before a previous flush instruction’s effects targeting the same line are completed, such an upcoming prefetch hint will be ignored by the CPU if the cache line is already in the cache, thereby reducing the actual activation rate.
Therefore, this highlights the importance of addressing the intrinsic disorder issue of prefetch instructions to ensure reliable and effective prefetch-based Rowhammer attacks.

\subsection{Combining Bank-Level Parallelism} \label{sec:rho_bank}

\begin{figure}[t!]
    % \hspace{-5mm}
    \centering
    \includegraphics[width=0.47\textwidth]{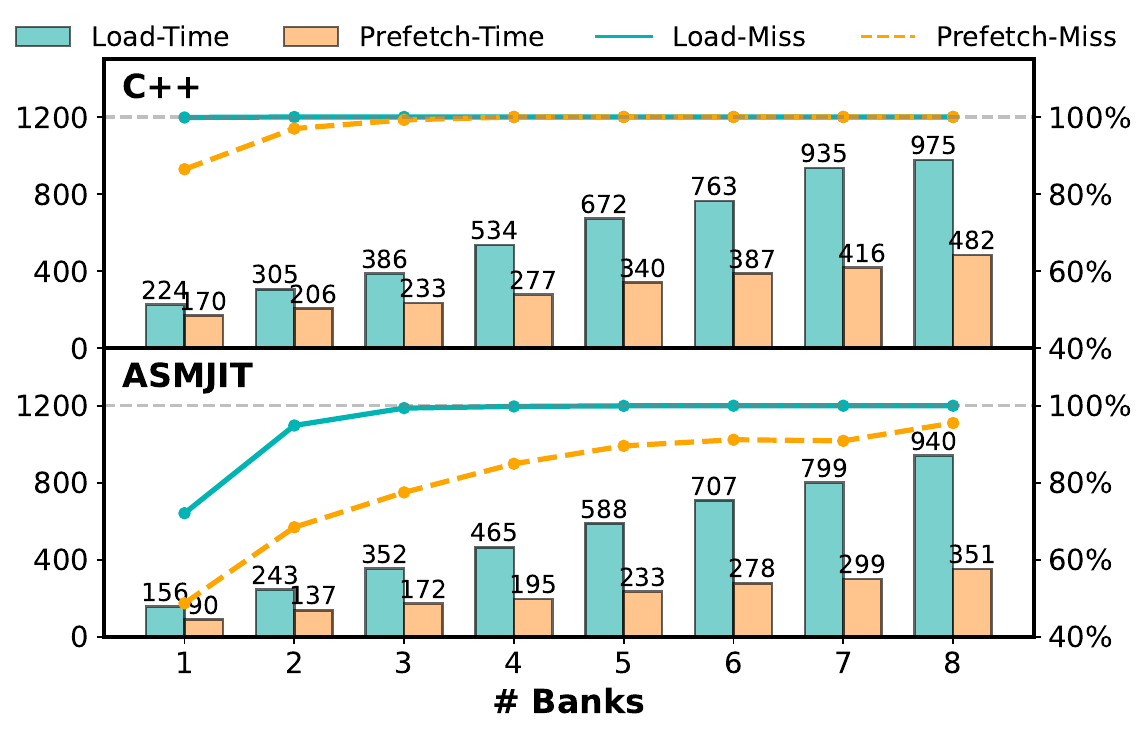}
    \caption{Average cache miss rate and attack time in \textit{ms} using the C++/AsmJit primitives with load/prefetch-based hammering on Comet Lake.} \label{fig:cachemiss}
    \vspace{-5mm}
\end{figure}

In order to further boost the activation rate, we introduce another critical multi-bank technique following a quite recent work~\cite{sledgehammer} and investigate its interplay with prefetch-based hammering.
% Another quite recent work, SledgeHammer~\cite{sledgehammer}, proposes the \textit{multi-bank hammering} technique that exploits bank-level parallelism in DRAMs to achieve faster RowHammer attacks.
Essentially, this technique exploits bank-level parallelism to share all activations across multiple banks (i.e., sacrificing the per-bank activation rate) and thus maximize the overall hammering throughput.
% Therefore, in this section, we investigate the impact of the multi-bank technique on prefetch-based hammering,
Using the same setup as Figure~\ref{fig:time}, we perform load/prefetch-based attacks with C++ and AsmJit primitives, ranging from 1 to 8 banks, and measure the average execution time and cache miss rate of each pattern.
We show the example results on Comet Lake in Figure~\ref{fig:cachemiss}.

Reflecting the previous section, the overall trend in cache miss rate further justifies that prefetching exhibits more severe disorder than loads.
Interestingly, besides increased parallelism, we observe another overlooked effect of multi-bank hammering that mitigates the disorder issue, as the cache miss rate increases with more banks.
One primary cause for such an effect is that the latency between hammer attempts on each bank is extended due to the interleaving of hammer attempts on other banks, which gradually reduces the degree of disorder and leads to an increasingly controlled hammering execution.
\label{reb:F-Q3-2}
As for hammering speed, our results demonstrate that prefetching doubles the attack speed when cache miss rates reach their peak.
This suggests that contrary to the conclusions drawn by~\cite{zenhammer}, even excluding the cache hit influence, prefetch instructions can still significantly enhance the hammering throughput.
% Additionally, by comparing the results between C++ and AsmJit primitives, we observe that cache miss rates can converge to 100\% more quickly using the former, whereas 
Additionally, our comparison between C++ and AsmJit primitives reveals that the former achieves full cache miss saturation much faster.
In contrast, the latter's cache miss rate remains well below 100\% even under 8 banks, but excessively high bank count would severely degrade per-bank activation rates.
Therefore, $\rho$Hammer adopts the preferable C++ primitive for prefetch-based hammering to ensure improved attack performance and reliability.

\begin{figure}[ht]
    \hspace{-7mm}
    \centering
    \includegraphics[width=0.47\textwidth]{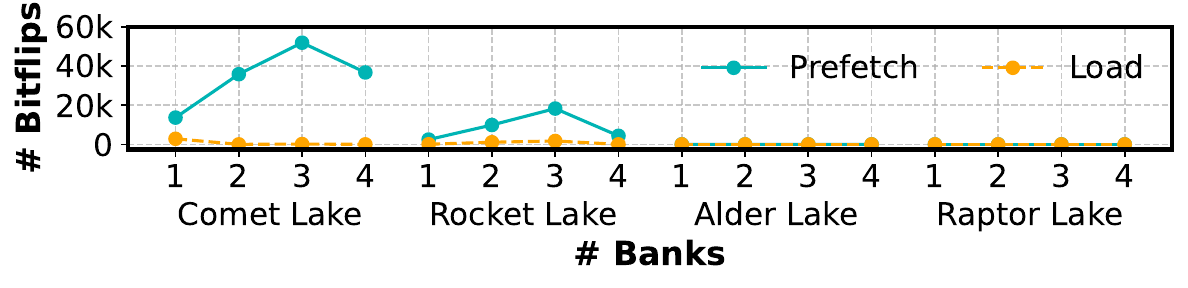}
    \caption{Overall effectiveness using load/prefetch-based hammering across 1-4 banks on all four architectures.} \label{fig:bank}
    \vspace{-5mm}
\end{figure}

To further demonstrate the advantages of multi-bank prefetch-based hammering, we measure the total number of bit flips during 2-hour load/prefetch-based fuzzing operations using varying numbers of banks across all four architectures.
Figure~\ref{fig:bank} evidently proves that the attack performance of prefetch-based hammering is significantly superior to load-based on Comet/Rocket Lake, further emphasizing the outstanding attack capabilities derived from prefetching.
Meanwhile, the results for Comet Lake align with our findings in Figure~\ref{fig:cachemiss}: the highest number of bit flips occurs when the cache miss rate is reaching its peak (i.e., \#Banks = 3).
After reaching the most flips (51,911), increasing the number of banks further would result in a gradual decline in hammering effectiveness, as expected.
Although the highest flip count here in the figure is at \#Banks = 3, it is important to clarify that the optimal number of banks could vary across architectures and DIMMs.
Therefore, for simplicity, the multi-bank configurations mentioned in the rest of this paper refer to the optimal bank number identified via fuzzing.

In stark contrast, the results on Alder/Raptor Lake shown in Figure~\ref{fig:bank}, reveal that combining the multi-bank technique with prefetch-based hammering is still insufficient to incur bit flips on these architectures.
\label{reb:F-Q4-1}
In fact, beyond this specific result, we have seen a sharp decline in bit flips on Alder Lake and a complete absence of flips on Raptor Lake, despite using the same flippable DIMMs, TRR-bypassing access patterns, and physical locations with identical extent of hammering tolerance.
These consistent failures strongly suggest that the more aggressive speculative execution mechanisms are the primary culprit suppressing our attack towards these newer architectures and more advanced hammering techniques are required to overcome their interference.

\subsection{Counter-Speculation $\rho$Hammering} \label{sec:rho_counter}

In this section, we elaborate $\rho$Hammer's counter-speculation technique that enables Rowhammer attacks on the latest architectures with intensified speculative disorder issues.
Generally, the technique combines runtime control flow obfuscation and intricate \texttt{NOP}-based pseudo-barriers, carefully inserting runtime latencies to counteract the adverse effects of branch prediction and OoO execution, while minimizing performance overhead to preserve the high throughput of prefetch-based hammering.

\textbf{Misleading the Branch Predictor.}
Firstly, we incorporate a control flow obfuscation technique to suppress the branch prediction's impact across loop iterations shown in Listing~\ref{lst:cpp}.
Specifically, we generate randomness from the \texttt{rdrand} and \texttt{rdtscp} instructions, which dynamically derives the indexes of multiple execution paths through additional bit-wise and arithmetic manipulations.
The obfuscation renders two pivotal components for branch prediction ineffective: 1) the branch target buffer (BTB), which stores the target addresses of recently encountered branches, are frequently invalidated; 2) the pattern history table (PHT), which relies on past branch outcomes to predict future branches, becomes barely reliable as it struggles to adapt to the fluctuating control flow that is only resolved at the beginning of each loop iteration.
By continually confusing the branch predictor, this technique forces the processor to fall back on a more conservative, in-order execution path.

\textbf{NOP-based Pseudo-barriers.}
We introduce another key technique of $\rho$Hammer to mitigate the OoO execution effects.
One of the critical hardware structures used for OoO execution is the reorder buffer (ROB), which tracks the original program order of instructions and ensures their results are committed correctly.
The ROB has a finite size, and once it is highly occupied, the processor’s ability to reorder instructions is constrained.
Therefore, we utilize the \texttt{NOP} instruction to consume reorder buffer slots while minimizing extra computational overhead.
After injecting sufficient \texttt{NOPs} between two subsequent instructions to saturate the ROB, their execution would be fully serialized~\cite{evict_spec_time}.
Building upon this concept, we build a pseudo-barrier strategy that inserts \texttt{NOPs} moderately but not excessively as we need to strike a balance between the hammering reliability and extra barrier overhead rather than ensure deterministic in-order behavior.  

\begin{figure}[t!]
    \centering
    \includegraphics[width=0.47\textwidth]{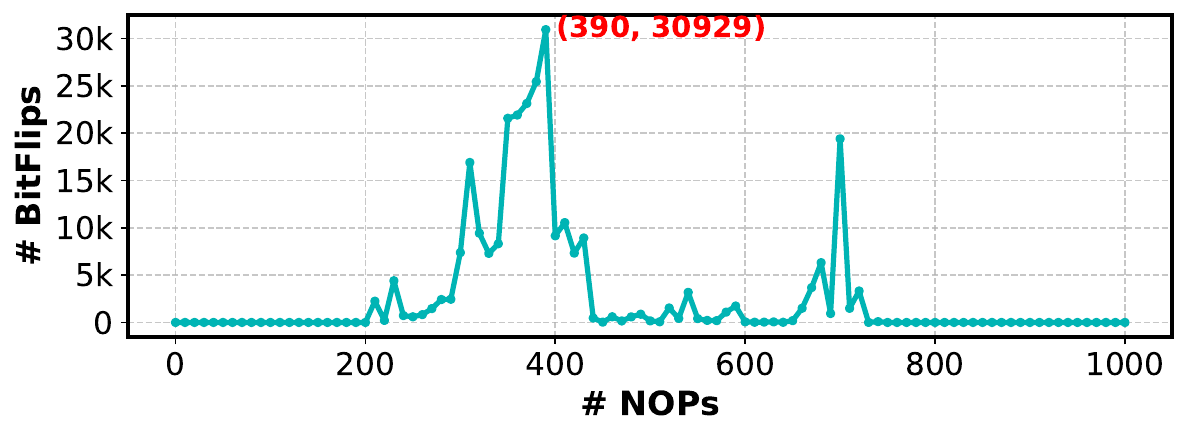}
    \caption{Number of bit flips with different \texttt{nop} counts. Each data point is the sum of sweeping the best pattern on Raptor Lake for 1,000 base addresses * 5 million activations.} \label{fig:nops}
    \vspace{-7mm}
\end{figure}

\textbf{Optimal NOP Count.}
% After applying the two counter-speculation techniques, we are begining to retrieve a few effective patterns and bit flips on the two latest architectures.
% Specifically, we insert numerous \texttt{nops} between each DRAM access and the branch obfuscation at the start of each pattern's loop.
We have observed that there is substantial room for improvement by discovering an optimal \texttt{NOP} range that strikes the perfect of such a balance.
As illustrated in Figure~\ref{fig:nops}, we show the example of sweeping one best pattern on Raptor Lake, where excessively low \texttt{NOP} count is insufficient to counter the OoO execution and the opposite sacrifices too much activation rate, both of which fail to yield any flip.
The optimum flip count (390) lies in the intermediate positive range, implying that a real attacker can take the same approach to search the optimum before exploitation.
In fact, the optimal \texttt{NOP} count is platform specific due to different OoO designs and ROB sizes, but one optimum is usually also effective across various patterns on the same platform (i.e., within the positive range).
Therefore, $\rho$Hammer incorporates a tuning phase to identify and use the optimal number of \texttt{NOP} instructions inserted between prefetches.
\label{reb:B-Q2-1}
However, when we attempt to adapt our counter-speculation technique to load-based hammering, we still cannot find any bit flip after trying all possible numbers of \texttt{NOPs} within the same [0, 1000] range on Raptor Lake, which also reflects the key factor leading to such a difference lies in the activation rate.
% We note that all hereinafter $\rho$Hammer experiments use the optimal \texttt{NOP} count by default.

\textbf{Pitfalls of Using Barrier Instructions.}
Prior work~\cite{zenhammer} studied the potential hazard of speculative execution and highlighted the selective use of the x86 fence instructions (i.e., \texttt{[L|S|M]FENCE}) for load-based hammering.
However, on the two newer architectures, we have observed zero bit flips regardless of how long we perform load-based hammering, both with and without fence instructions.
As discussed earlier, prefetching is inherently more susceptible to speculative disorder, so we further investigate whether existing fence strategies are useful for prefetching and compare them with our \texttt{NOP}-based pseudo-barrier technique.

\begin{normalsize}
\begin{table}[h]
    \vspace{-3mm}
    \setlength{\extrarowheight}{2pt}
    \setlength{\tabcolsep}{6pt} 
    \centering
    \small
    \begin{tabular}{ccccccc}
        \bottomrule
        \textbf{Arch.} & \textbf{None} & \textbf{\texttt{CPUID}} & \textbf{\texttt{MFENCE}} & \makecell{\textbf{\texttt{LFENCE}}\\(load)} & \makecell{\textbf{\texttt{LFENCE}}\\(prefetch)} & \textbf{\texttt{NOP}} \\
        \hline
        Alder & 0 & 0 & 0 & 0 & 8,113 & 5,250 \\
        Lake & 93.7 & 2,104.4 & 1,179.2 & 649.6 & 211.7 & 212.6 \\
        \hline
        Raptor & 0 & 0 & 0 & 0 & 1,903 & 7,210 \\
        Lake & 92.8 & 1,861.3 & 1,149.6 & 466.2 & 210.4 & 212.4 \\
        \bottomrule
    \end{tabular}
    \caption{Comparison of different barriers on Alder Lake and Raptor Lake. Upper digits are number of bit flips and lower ones are time in \textit{ms}.}
    \label{tab:barriers}
    \vspace{-7mm}
\end{table}
\end{normalsize}

Specifically, we perform prefetch-based sweeping over best patterns for 200 base addresses * 10 million activations using various barrier instructions as well as our \texttt{NOP}-based pseudo-barriers, all with control flow obfuscation enabled.
The maximal number of bit flips and the completion time are recorded in Table~\ref{tab:barriers}.
The Intel Software Developer's Manual~\cite{intel_soft_dev} states that \texttt{PREFETCH\textit{h}} instructions are not ordered with respect to the fence instructions and other \texttt{PREFETCH\textit{h}} instructions, but are ordered with respect to serializing instructions such as \texttt{CPUID}.
However, the table indicates that using \texttt{CPUID} for serializing prefetches and \texttt{MFENCE} is too time-consuming to meet the minimum activation rate, which is even worse than the conventional load+\texttt{LFENCE} strategy.

From the table, it seems that \texttt{LFENCE} is also effective for prefetch-based hammering, exhibiting execution time comparable to that of the \texttt{NOP}-based pseudo-barriers, especially on Alder Lake.
Although this appears to contradict the manual's statement that prefetches are not ordered by fences, we argue that the manual remains correct in this regard.
To further investigate, we reuse the previous AsmJit primitive that leverages immediate rather than indirect addressing (i.e., no data dependency chain), but with \texttt{LFENCE} inserted.
Under this modified setup, we observe a sharp decline in the bit flip count, confirming that the effectiveness of \texttt{LFENCE} does not result from \emph{direct ordering} on prefetches.
Instead, the effect stems from \texttt{LFENCE} impacting the address calculation process at line 4\&5 of Listing~\ref{lst:cpp}, where the target address of the current prefetch is explicit memory read that must be architecturally resolved before issuing the next, thereby imposing an \emph{indirect ordering} on prefetches.

\begin{table*}[t!]
    \setlength{\tabcolsep}{6pt}
    \centering
    \small
    
    \renewcommand\arraystretch{1.5}
    \begin{tabular}{c|m{0.27\textwidth}|m{0.27\textwidth}|m{0.27\textwidth}}
        \bottomrule
        \multirow{2}{*}{\textbf{Arch.}} & \multicolumn{3}{c}{\textbf{DRAM Geometry (Size, \# Rank, \# Bank)}} \\
        \cline{2-4}
        & \centering (8G, 1, 16) & \centering (16G, 2, 16) & \centering (32G, 2, 16) \tabularnewline
        \hline
        \makecell{\textbf{Comet Lake} \\ \textbf{Rocket Lake}} &
        \textbf{Bank Func:} (16, 19), (15, 18), (14, 17), (6, 13); \textbf{Row:} 17--32 &
        \textbf{Bank Func:} (17, 21), (16, 20), (15, 19), (14, 18), (6, 13); \textbf{Row:} 18--33 &
        \textbf{Bank Func:} (17, 21), (16, 20), (15, 19), (14, 18), (6, 13); \textbf{Row:} 18--34 \\
        \hline
        \makecell{\textbf{Alder Lake} \\ \textbf{Raptor Lake}} &
        \textbf{Bank Func:} (14, 17, 21, 26, 29, 32), (15, 18, 20, 23, 24, 27, 30), (16, 19, 22, 25, 28, 31), (9, 11, 13); \textbf{Row:} 17--32 &
        \textbf{Bank Func:} (14, 18, 26, 29, 32), (16, 20, 23, 24, 27, 30, 33), (17, 21, 22, 25, 28, 31), (15, 19), (9, 11, 13); \textbf{Row:} 18--33 &
        \textbf{Bank Func:} (14, 18, 26, 29, 32), (16, 20, 23, 24, 27, 30, 33), (17, 21, 22, 25, 28, 31, 34), (15, 19), (9, 11, 13); \textbf{Row:} 18--34 \\
        \bottomrule
    \end{tabular}
    
    \caption{Reverse-engineered DRAM address mapping on the four most recent Intel architectures. All configurations are single-channel, single-DIMM, with different DRAM geometry.}
    \label{tab:re_res}
    \vspace{-7mm}
\end{table*}

\subsection{Summary \& Implications} \label{sec:rho_sum}
Here we summarize our observations and elaborate on why $\rho$Hammer is fundamentally more preferable than load-based methods.

In order to achieve higher activation rates, prior efforts have demonstrated using multi-threaded hammering~\cite{Throwhammer,SGX-Bomb}, mainly on DDR3 devices.
However, as reported by a quite recent work~\cite{WhistleBlower}: on DDR4 devices equipped with TRR, multi-threaded hammering is in fact much less effective than single-threaded and the trend could be even worse as the growing number of threads.
The root cause is that effective bypassing the TRR requires a strict access order (i.e., the non-uniform pattern used in this paper).
Once multiple threads hammer concurrently, their asynchronous requests collide in the memory-controller queue, disturb the pattern, and trigger extra refreshes by the TRR sampler, while enforcing a global order with synchronization mechanisms such as locks or semaphores re-introduces serialization and further degrade the activation rate to even lower than just using a single thread~\cite{WhistleBlower}.

In the case of single-threaded hammering, recall in Figure~\ref{fig:cachemiss}, under a full cache-miss situation, tight loops of prefetch-based hammering are significantly faster than load-based counterparts while targeting same aggressors.
Moreover, the former achieves evidently higher overall hammering effectiveness and flip rate, which we will soon demonstrate in Section~\ref{sec:eval}.
These results give two important implications:
1) \textbf{Single-threaded loads cannot saturate the DRAM bandwidth}, a widely acknowledged fact that is also supported by the massive reported results through public micro-benchmarks such as Stream~\cite{stream}.
2) \textbf{Prefetches better utilize the single-threaded DRAM bandwidth than loads}.
The underlying $\mu$arch reason is that a regular load occupies a load-queue entry until the line returns, throttling the issue rate once these structures are filled.
By contrast, as previously stated, a prefetch instruction does not stall and is immediately \textit{marked as complete} in the ROB as soon as the address translation is done~\cite{intel_arch_opt}.
In support of this, our results in Table~\ref{tab:barriers} have shown prefetch+\texttt{LFENCE} is much faster than load+\texttt{LFENCE} as \texttt{LFENCE} only orders pending load requests that are \textit{not marked as complete}, also noted by~\cite{amd_prefetch} on AMD platforms.

\section{Evaluation} \label{sec:eval}

\subsection{DRAM Address Mapping Recovery} \label{sec:eval_re}

Table~\ref{tab:re_res} summarizes our reverse-engineering results across all tested setups.
We observe that, on a given architecture, the DRAM address mappings are consistent across DIMMs with identical geometry, with Comet/Rocket Lake sharing one mapping scheme and Alder/Raptor Lake adopting another.
Then, we compare our method against all publicly available existing tools~\cite{drama,dramdig,zenhammer} (properly configured to our setups) and record the average timing results of 50 independent runs each in Table~\ref{tab:re_time}.
We only observe that DRAMDig\footnote{\url{https://github.com/dramdig/DRAMDig}} and DARE\footnote{\url{https://github.com/comsec-group/zenhammer/tree/dare}} are able to yield correct results on Comet/Rocket Lake.
Specifically, our method outperforms DRAMDig by more than two orders of magnitude in terms of runtime speed.
As for DARE, even when we allocate the maximum number of superpages for bank function coloring, the results still exhibited quite a few errors (accuracy: 34/50 on Comet Lake, 39/50 on Raptor Lake).
Furthermore, none of the prior methods are adaptable to the latest Alder/Raptor Lake, where both DRAMA\footnote{\url{https://github.com/isec-tugraz/drama}} and DARE fail to produce any correct mappings, while DRAMDig terminates prematurely due to the requirement of pure row bits.
These findings highlight the superior efficiency and scalability of our reverse-engineering technique across recent architectures.

\begin{table}[h]
    \setlength{\tabcolsep}{2.5pt}
    \centering
    \small
    
    \renewcommand\arraystretch{1.4}
    \begin{tabular}{cm{1.5cm}m{1.5cm}m{1.5cm}m{1.5cm}}
        \bottomrule
        \textbf{} & \centering\textbf{i7-10700K} & \centering\textbf{i7-11700} & \centering\textbf{i9-12900} & \centering\textbf{i7-14700K} \tabularnewline
        \hline
        \textbf{DRAMA} & \centering - & \centering - & \centering - & \centering - \tabularnewline
        \hline
        \textbf{DRAMDig} & \centering 867.6s & \centering 1,329.9s & \centering - & \centering - \tabularnewline
        \hline
        \textbf{DARE} & \centering 36.5s* & \centering 33.1s* & \centering - & \centering - \tabularnewline
        \hline
        \textbf{$\rho$Hammer} & \centering 8.5s & \centering 6.1s & \centering 4.6s & \centering 4.1s \tabularnewline
        \bottomrule
    \end{tabular}
    
    \caption{Reverse-engineering time comparing to prior art~\cite{drama,dramdig,zenhammer}. (*) denotes partially non-deterministic while (-) denotes no correct result or the process aborts with failure.}
    \label{tab:re_time}
    \vspace{-7mm}
\end{table}
% \addtocounter{footnote}{-3}
% \footnotetext[1]{\phantomsection\label{fn:drama}\url{https://github.com/isec-tugraz/drama}}
% \footnotetext[2]{\phantomsection\label{fn:dramdig}\url{https://github.com/dramdig/DRAMDig}}
% \footnotetext[3]{\phantomsection\label{fn:dare}\url{https://github.com/comsec-group/zenhammer/tree/dare}}

\subsection{Overall Hammering Effectiveness} \label{sec:eval_fuzz}

\begin{table*}[t!]
    \renewcommand{\arraystretch}{1.3}
    \small
    \centering
    \begin{minipage}{\textwidth}
    \centering
    \begin{tabular}{cm{1.7cm}m{1.7cm}m{1.7cm}m{1.75cm}m{1.7cm}m{1.7cm}m{1.7cm}m{1.7cm}}
        \bottomrule
        \multicolumn{1}{c|}{\multirow{2}{*}{\textbf{DIMM}}} & \multicolumn{4}{c|}{\textbf{Comet Lake}} & \multicolumn{4}{c}{\textbf{Rocket Lake}} \\
        \cline{2-9}
        & \multicolumn{1}{|c|}{BL-S} & \multicolumn{1}{c|}{BL-M} & \multicolumn{1}{c|}{$\rho$-S} & \multicolumn{1}{c|}{$\rho$-M} & \multicolumn{1}{c|}{BL-S} & \multicolumn{1}{c|}{BL-M} & \multicolumn{1}{c|}{$\rho$-S} & \multicolumn{1}{c}{$\rho$-M} \\
        \hline
        \multicolumn{1}{c}{\cellcolor{gray!20}$\mathcal{S}_1$} & \multicolumn{1}{c}{\cellcolor{red!20}2840, 529} & \multicolumn{1}{c}{\cellcolor{red!20}0, 0} & \multicolumn{1}{c}{\cellcolor{red!20}9117, 1506} & \multicolumn{1}{c}{\cellcolor{red!20}121431, 25361} & \multicolumn{1}{c}{\cellcolor{cyan!20}58, 18} & \multicolumn{1}{c}{\cellcolor{cyan!20}1088, 490} & \multicolumn{1}{c}{\cellcolor{cyan!20}10458, 2547} & \multicolumn{1}{c}{\cellcolor{cyan!20}24548, 5257} \\
        $\mathcal{S}_2$ & \makecell[c]{12288, 2274} & \makecell[c]{437, 178} & \makecell[c]{55270, 9720} & \makecell[c]{87824, 24177} & \makecell[c]{3815, 720} & \makecell[c]{7448, 1255} & \makecell[c]{17200, 1254} & \makecell[c]{42485, 4831} \\
        \multicolumn{1}{c}{\cellcolor{gray!20}$\mathcal{S}_3$} & \multicolumn{1}{c}{\cellcolor{red!20}36406, 4439} & \multicolumn{1}{c}{\cellcolor{red!20}9373, 521} & \multicolumn{1}{c}{\cellcolor{red!20}67552, 1566} & \multicolumn{1}{c}{\cellcolor{red!20}205742, 7526} & \multicolumn{1}{c}{\cellcolor{cyan!20}9772, 3041} & \multicolumn{1}{c}{\cellcolor{cyan!20}10211, 923} & \multicolumn{1}{c}{\cellcolor{cyan!20}48419, 1235} & \multicolumn{1}{c}{\cellcolor{cyan!20}\textbf{94395, 2896}} \\
        $\mathcal{S}_4$ & \makecell[c]{50700, 7839} & \makecell[c]{1472, 199} & \makecell[c]{76645, 5223} & \makecell[c]{\textbf{257881, 13880}} & \makecell[c]{6560, 2030} & \makecell[c]{4323, 1103} & \makecell[c]{53706, 3334} & \makecell[c]{81983, 5824} \\
        \multicolumn{1}{c}{\cellcolor{gray!20}$\mathcal{S}_5$} & \multicolumn{1}{c}{\cellcolor{red!20}365, 28} & \multicolumn{1}{c}{\cellcolor{red!20}130, 38} & \multicolumn{1}{c}{\cellcolor{red!20}722, 123} & \multicolumn{1}{c}{\cellcolor{red!20}3660, 446} & \multicolumn{1}{c}{\cellcolor{cyan!20}79, 12} & \multicolumn{1}{c}{\cellcolor{cyan!20}171, 14} & \multicolumn{1}{c}{\cellcolor{cyan!20}301, 12} & \multicolumn{1}{c}{\cellcolor{cyan!20}963, 80} \\
        $\mathcal{H}_1$ & \makecell[c]{136, 18} & \makecell[c]{33, 19} & \makecell[c]{406, 30} & \makecell[c]{1714, 441} & \makecell[c]{0, 0} & \makecell[c]{23, 4} & \makecell[c]{918, 68} & \makecell[c]{1498, 271} \\
        \multicolumn{1}{c}{\cellcolor{gray!20}$\mathcal{M}_1$} & \multicolumn{1}{c}{\cellcolor{red!20}0, 0} & \multicolumn{1}{c}{\cellcolor{red!20}0, 0} & \multicolumn{1}{c}{\cellcolor{red!20}0, 0} & \multicolumn{1}{c}{\cellcolor{red!20}0, 0} & \multicolumn{1}{c}{\cellcolor{cyan!20}0, 0} & \multicolumn{1}{c}{\cellcolor{cyan!20}0, 0} & \multicolumn{1}{c}{\cellcolor{cyan!20}0, 0} & \multicolumn{1}{c}{\cellcolor{cyan!20}0, 0} \\
        \bottomrule
    \end{tabular}
    \end{minipage}

    \vspace{2mm}
    
    \begin{minipage}{\textwidth}
    \centering
    \begin{tabular}{cm{1.7cm}m{1.7cm}m{1.7cm}m{1.75cm}m{1.7cm}m{1.7cm}m{1.7cm}m{1.7cm}}
        \bottomrule
        \multicolumn{1}{c|}{\multirow{2}{*}{\textbf{DIMM}}} & \multicolumn{4}{c|}{\textbf{Alder Lake}} & \multicolumn{4}{c}{\textbf{Raptor Lake}} \\
        \cline{2-9}
        & \multicolumn{1}{|c|}{BL-S} & \multicolumn{1}{c|}{BL-M} & \multicolumn{1}{c|}{$\rho$-S} & \multicolumn{1}{c|}{$\rho$-M} & \multicolumn{1}{c|}{BL-S} & \multicolumn{1}{c|}{BL-M} & \multicolumn{1}{c|}{$\rho$-S} & \multicolumn{1}{c}{$\rho$-M} \\
        \hline
        \multicolumn{1}{c}{\cellcolor{gray!20}$\mathcal{S}_1$} & \multicolumn{1}{c}{\cellcolor{blue!20}0, 0} & \multicolumn{1}{c}{\cellcolor{blue!20}0, 0} & \multicolumn{1}{c}{\cellcolor{blue!20}169, 67} & \multicolumn{1}{c}{\cellcolor{blue!20}571, 158} & \multicolumn{1}{c}{\cellcolor{orange!20}0, 0} & \multicolumn{1}{c}{\cellcolor{orange!20}0, 0} & \multicolumn{1}{c}{\cellcolor{orange!20}20, 6} & \multicolumn{1}{c}{\cellcolor{orange!20}266, 53} \\
        $\mathcal{S}_2$ & \makecell[c]{3, 1} & \makecell[c]{2, 1} & \makecell[c]{582, 137} & \makecell[c]{\textbf{918, 233}} & \makecell[c]{0, 0} & \makecell[c]{1, 1} & \makecell[c]{490, 137} & \makecell[c]{\textbf{2113, 1564}} \\
        \multicolumn{1}{c}{\cellcolor{gray!20}$\mathcal{S}_3$} & \multicolumn{1}{c}{\cellcolor{blue!20}2, 2} & \multicolumn{1}{c}{\cellcolor{blue!20}10, 7} & \multicolumn{1}{c}{\cellcolor{blue!20}107, 31} & \multicolumn{1}{c}{\cellcolor{blue!20}696, 141} & \multicolumn{1}{c}{\cellcolor{orange!20}8, 7} & \multicolumn{1}{c}{\cellcolor{orange!20}0, 0} & \multicolumn{1}{c}{\cellcolor{orange!20}183, 53} & \multicolumn{1}{c}{\cellcolor{orange!20}924, 140} \\
        $\mathcal{S}_4$ & \makecell[c]{52, 48} & \makecell[c]{4, 2} & \makecell[c]{127, 12} & \makecell[c]{413, 89} & \makecell[c]{0, 0} & \makecell[c]{4, 4} & \makecell[c]{62, 22} & \makecell[c]{536, 99} \\
        \multicolumn{1}{c}{\cellcolor{gray!20}$\mathcal{S}_5$} & \multicolumn{1}{c}{\cellcolor{blue!20}0, 0} & \multicolumn{1}{c}{\cellcolor{blue!20}0, 0} & \multicolumn{1}{c}{\cellcolor{blue!20}7, 2} & \multicolumn{1}{c}{\cellcolor{blue!20}22, 7} & \multicolumn{1}{c}{\cellcolor{orange!20}0, 0} & \multicolumn{1}{c}{\cellcolor{orange!20}0, 0} & \multicolumn{1}{c}{\cellcolor{orange!20}18, 4} & \multicolumn{1}{c}{\cellcolor{orange!20}20, 6} \\
        $\mathcal{H}_1$ & \makecell[c]{0, 0} & \makecell[c]{0, 0} & \makecell[c]{40, 176} & \makecell[c]{402, 27} & \makecell[c]{0, 0} & \makecell[c]{0, 0} & \makecell[c]{92, 6} & \makecell[c]{384, 34} \\
        \multicolumn{1}{c}{\cellcolor{gray!20}$\mathcal{M}_1$} & \multicolumn{1}{c}{\cellcolor{blue!20}0, 0} & \multicolumn{1}{c}{\cellcolor{blue!20}0, 0} & \multicolumn{1}{c}{\cellcolor{blue!20}0, 0} & \multicolumn{1}{c}{\cellcolor{blue!20}0, 0} & \multicolumn{1}{c}{\cellcolor{orange!20}0, 0} & \multicolumn{1}{c}{\cellcolor{orange!20}0, 0} & \multicolumn{1}{c}{\cellcolor{orange!20}0, 0} & \multicolumn{1}{c}{\cellcolor{orange!20}0, 0} \\
        \hline
    \end{tabular}
    \end{minipage}
    
    \caption{Bit flip counts (total, best pattern) during 2-hour fuzzing on all platforms, for each of which we record the result using baseline/$\rho$Hammer (BL/$\rho$) and single-bank/multi-bank (S/M).}
    \vspace{-8mm}
    \label{tab:eval_fuzz}
\end{table*}

We implement $\rho$Hammer by extending the code from the existing non-uniform hammering tools, namely Blacksmith\footnote{\url{https://github.com/comsec-group/blacksmith}} and ZenHammer\footnote{\url{https://github.com/comsec-group/zenhammer}}.
In our attack evaluation, we use their original load-based non-uniform hammering as the baseline and report the performance of both $\rho$Hammer and the baseline under single-bank and optimal multi-bank configurations to establish fair comparisons.

Table~\ref{tab:eval_fuzz} summarizes the bit flip counts from all effective patterns and the best pattern during our large-scale 2-hour fuzzing operations for all DIMMs and architectures.
Across the board, $\rho$Hammer significantly outperforms the baseline in raw flip counts, which is most evident in the following per-architecture standout results (marked in bold):
For example, on Comet Lake - $\mathcal{S}_4$, $\rho$-M achieves 257,881 total bit flips, far surpassing the 50,700 from BL-S and making it the most flip-prone across all configurations; while on Raptor Lake - $\mathcal{S}_2$ where the baseline produces negligible flips, $\rho$-M achieves as many as 2,113.

Generally, the results demonstrate that $\rho$Hammer not only significantly amplifies the Rowhammer threat on architectures where baseline attacks remain limitedly effective (e.g., Comet/Rocket Lake), but more importantly, it revives Rowhammer on the most recent platforms like Alder/Raptor Lake, where baseline techniques almost completely fail to induce any meaningful flips.
Moreover, $\rho$-M always outperforms $\rho$-S across all setups, which highlights the crucial role of the multi-bank technique in maximizing the full potential of prefetch-based hammering.
In contrast, the baseline configurations display a number of counterintuitive cases where BL-S outperforms BL-M, especially on Comet Lake.
This aligns with the fundamental limitation of loads as discussed in Section~\ref{sec:rho_sum}: since load-based hammering is heavily constrained by activation rate, adding more banks can exacerbate this bottleneck rather than alleviate it.
Additionally, on platforms with relatively weaker speculative disorder (as primarily observed on Comet Lake; recall Figure~\ref{fig:cachemiss}), the disorder effect from single-bank load-based hammering may already be minimal, thus little benefit from alleviated disorder could be gained by distributing accesses across multiple banks.

\subsection{Practical Exploitability} \label{sec:eval_sweep}

To more intuitively justify $\rho$Hammer's security impact, we further: 1) analyze flip rates that reflect the speed and chance of finding useful flips; 2) perform end-to-end PTE-corruption attack POCs.

\begin{figure}[h]
    \vspace{-5mm}
    \centering
    \includegraphics[width=0.47\textwidth]{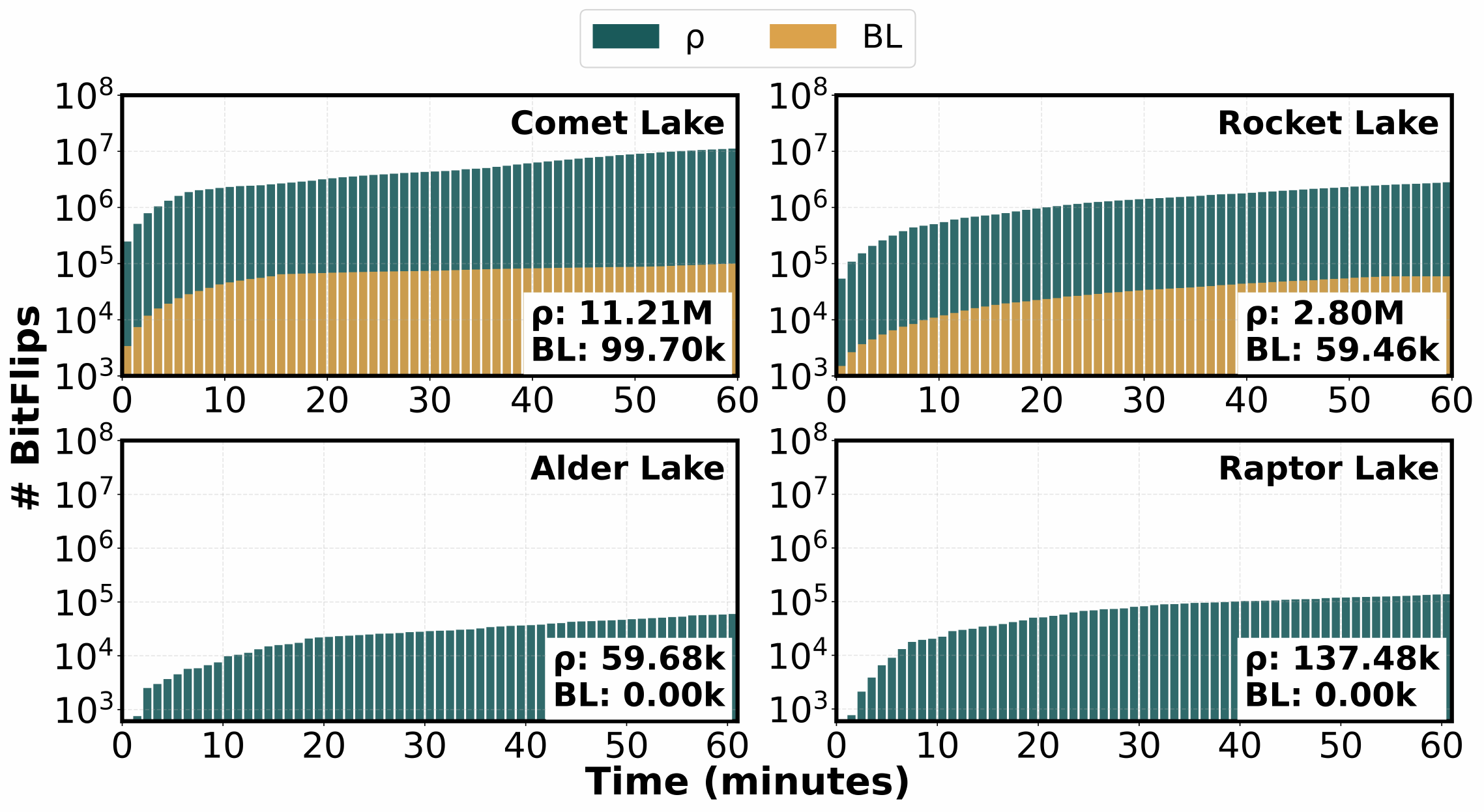}
    \caption{Accumulative average number of bit flips over iterative 1-hour sweeping on the four architectures.} \label{fig:eval_sweep}
    \vspace{-5mm}
\end{figure}

% To justify $\rho$Hammer's contribution to memory templating in real attacks, as mentioned in the threat model, we further analyze flipping rates over long sweeping processes, the higher of which inherently promote the chance of finding surgical bit positions that would be later used for exploitation (e.g., physical address range within a page table entry~\cite{rubicon}).
% While fuzzing demonstrates the overall effectiveness of a specific hammering strategy by indicating the difficulty of identifying effective patterns, we further validate $\rho$Hammer's practicality and reproducibility in real attacks by analyzing bit flip rates over long sweeping processes.
\textbf{Flip Rate.}
Figure~\ref{fig:eval_sweep} illustrates the increasing trends of bit flips during 1-hour sweeping using \emph{the best pattern} and \emph{the better of single/multi-bank} respectively for $\rho$Hammer and baselines.
% To ensure fair comparisons, we select the $\mathcal{S}_1$ DIMM, where both $\rho$Hammer and baselines are effective, and use their better of single/multi-bank.
Given that Rowhammer vulnerability depends on the DIMM location, we perform sweeping across four architectures using the same set of non-repeating row addresses.
For the Alder/Raptor Lake where the baseline produces zero flips, we use $\rho$Hammer's best pattern for sweeping the baseline as a fallback.

The results reveal striking differences: on Comet/Rocket Lake, $\rho$Hammer achieves an average flip rate of 187K/min and 47K/min, significantly faster than the baseline by 112.4x and 47.1x.
Moreover, on Alder/Raptor Lake, the baselines fail to reproduce any flips even if we switch to other DIMMs (e.g., $\mathcal{S}_4$) and patterns with occasional flips during fuzzing, but $\rho$Hammer still reaches the impressive average flip rates of 995/min and 2,291/min, which are also much higher compared with previous data (e.g., $\sim$133/min on Alder Lake reported by~\cite{sledgehammer}).
% These flip rates are far more than sufficient for practical attacks when referred to previous data (e.g., $\sim$133/min on Alder Lake reported by~\cite{sledgehammer}).
Another noteworthy finding is that $\rho$Hammer's bit flips smoothly progress over time (remind that the figure is logarithmic-scaled), which indicates that desired flips could be equally found at all positions.

\textbf{End-to-End Attack.}
Combining the latest massaging techniques proposed by Rubicon~\cite{rubicon}, we implement and perform end-to-end PTE corruption attacks on the newer two architectures.
We do not rely on superpages but sweep discrete 4MB contiguous memory regions and template all bit flip locations, which is the largest possible size an unprivileged attacker can manage to assure contiguity by exhausting the Linux's buddy allocator.
Using the best pattern, the templating phase on Alder/Raptor Lake platforms respectively identifies: 892/846 total flips within 38/35s, of which 61/43 are exploitable.
Here, \emph{exploitable} flips are those residing in a desired sub-range of PTE frame number (e.g., [12, 19]), while their physical location and flip direction allow self-reference after being massaged into a page table page.
Subsequently, we successfully achieve page table read/write capability on both platforms using an average runtime of only 3min 8s (Min: 54s; Max: 6min 44s) and 51s (Min: 20s; Max: 1min 55s), each for five independent trials.

\section{Discussion} \label{sec:dis}

\textbf{Mitigations.}
Beyond commodity TRR, another production-level defense proposed by Intel is the pseudo-TRR (pTRR)~\cite{intel_xeon}, which refreshes potential victim rows preemptively.
% One concurrent work, MCSEE~\cite{mcsee}, recently reported that Raptor Lake platforms have been integrated with pTRR as Rowhammer mitigation.
We identified an option named "Rowhammer Prevention" in the BIOS of our Alder/Raptor Lake platform (MSI motherboard model: PRO Z790-A WIFI DDR4).
% which, according to Intel's response, should be the pTRR support.
According to Intel, this option should be the support for the integrated pTRR mitigation on Alder/Raptor Lake CPUs.
When we enabled this option and repeated our hammering experiments, nearly all previously observed bit flips were eliminated.
We therefore advocate OEM support for such a feature on future Intel platforms and recommend that end users enable it as an effective mitigation against our attacks.
% Beyond commodity TRR, another production-level defense proposed by Intel is the pseudo-TRR~\cite{intel_xeon} (pTRR) that refreshes rows pre-emptively, yet it is only enabled when both the MC and the DIMM are compliant.
% Unfortunately, as reported in~\cite{trrespass} and to the best of our knowledge, the consumer-grade CPUs tested in this paper do not support this feature.
On the DRAM side, all our tested DIMMs set their Maximum Activation Count (MAC) value that indicates their claimed Rowhammer tolerance to "unlimited", usually derived from the less potent load-based hammering tests or even no thorough inspection, which is also noted by~\cite{trrespass,sledgehammer,utrr}.
However, as prefetch-based hammering exacerbates the threat, such bottom-line confidence is more misplaced than ever.
Recently in academia, numerous schemes have also been proposed, such as address-mapping scrambling that reorders the bank/row mapping using a boot-time key~\cite{cube}, and randomized row-swapping that periodically exchanges the contents of random row pairs to keep aggressor accesses from concentrating on the same victim~\cite{randomized_row_swap,scalable_row_swap,shadow}.
These efforts are expected to mitigate our attack because they would break the TRR-bypassing pattern and disperse the activations.
However, the additional remapping circuitry demands non-trivial modifications to both DRAM chips and memory controllers, which is unlikely to be widely deployed in the near term.

\textbf{Towards Future Research on DDR5.}
Recently prevailing DDR5 devices integrate more intricate defense mechanisms than TRR, such as refresh management (RFM)~\cite{jedec_ddr5}.
Thus, we have not observed any effective pattern on our setups with DDR5 DIMMs, which is also acknowledged by a concurrent non-uniform hammering work~\cite{posthammer}.
Nevertheless, our prefetch-based hammering paradigm would inspire future research on DDR5 by 1) supporting higher activation rate against doubled refresh rate, and 2) increasing the likelihood of multi-bit errors for on-die ECC corruption~\cite{jedec_ddr5,ECC}.
We have evaluated our reverse-engineering tool on Alder/Raptor Lake DDR5 systems and observed that further efforts are required to identify extra \emph{sub-channel} functions.
Another concurrent study~\cite{not_so_refreshing} shows that the RFM refreshing behavior forms an accurate new side-channel on GDDR6 and LPDDR5 devices.
Combining this new primitive with our efficient algorithm is thus an interesting future work that would further extend $\rho$Hammer towards DDR5.
%待加：bridge DDR5逆向

\section{Related Work} \label{sec:rw}

\textbf{DRAM Address Mapping Reverse-Engineering.}
As discussed in Section~\ref{sec:re_prelim}, Pessl et al.~\cite{drama} conducted the pioneering work that introduced a brute-force strategy for mapping recovery.
To reduce the brute-force search space, Wang et al.~\cite{dramdig} proposed a knowledge-assisted approach that excludes pure row bits before searching.
Jattke et al.~\cite{zenhammer} extended the strategy to AMD platforms with extra consideration of a constant address offset.
To the best of our knowledge, only two prior studies by Xiao et al.~\cite{one_Bit_Flips_One_Cloud_Flops} and Marazzi et al.~\cite{risc-h} exclude brute-forcing.
However, the former is limited to identifying simple single-bit functions on RISC-V platforms, while both this work and Wang et al.~\cite{dramdig} have found the latter fails to obtain correct mappings on more recent x86 platforms.

\textbf{Recent Advances in Rowhammer.}
Orosa et al.~\cite{deeper_look_into} conducted a systematic analysis of Rowhammer vulnerability across 272 chips and revealed that the problem may vary with physical locations, which is the reason for our controlling the base physical addresses for comparisons.
Frigo et al.~\cite{trrespass} proposed the many-sided hammering strategy that bypasses TRR by concealing the accesses of real aggressor rows using dummy ones.
Ridder et al.~\cite{smash} further introduced synchronized many-sided hammering from JavaScript, leveraging \texttt{NOPs} to align memory accesses with DRAM refreshes and bypass TRR's sampler, whose goal fundamentally differs from our use of \texttt{NOPs} as memory barriers.
Unlike previous uniform hammering approaches, Jattke et al.~\cite{blacksmith} presented the pioneering non-uniform hammering tool against TRR-protected DDR4 DIMMs, which could fuzz hammer patterns with aggressor pairs featuring various frequency-domain parameters.
Kang et al.~\cite{sledgehammer} first exploited bank-level parallelism to amplify Rowhammer effects.
Although they have observed the technique being the most effective on 6-10th generations with a maximum 13.1x increase on i7-6700, their total flip counts also suffered much reduction when targeting the 11/12th generations.
Jattke et al.~\cite{zenhammer} extended their non-uniform hammering approach to AMD Zen platforms.
They compared the activation rates of different types of DRAM accessing instructions, including \texttt{PREFETCHNTA}, but simply attributed the rate improvement to cache hits and suggested not to use such instructions as opposed to our work.
Zhang et al.~\cite{implicit_hammer,pthammer} introduced the concept of \emph{implicit} hammer attacks, where memory accesses are induced via system calls or page table walks to target addresses that the attacker cannot explicitly access.
Although prefetching results in implicit memory accesses, our attack remains fundamentally \emph{explicit}, as it must directly specify target addresses for prefetching.

\section{Conclusion}

In this paper, we revisit the viability of Rowhammer attacks on recent Intel architectures and identify key bottlenecks rendering load-based state-of-the-arts ineffective.
By introducing $\rho$Hammer, %we systematically address the mapping complexity, activation rate limitations, and speculative disorder issues that hinder attack feasibility.
we systematically address challenges related to DRAM address mapping complexity, limited activation rates, and speculative hammering disorder. %, all of which hinder attack feasibility. 
Our contribution involves a fast and accurate reverse-engineering method, a high-throughput prefetch-based hammering paradigm combining bank parallelism, and a counter-speculation technique tailored to maintain hammering order while maximizing the potential of prefetching.
Extensive evaluation shows that $\rho$Hammer not only amplifies attack effectiveness on known vulnerable platforms, but also enables practical exploitation on newer ones such as Intel Raptor Lake for the first time, calling for a renewed consideration of Rowhammer on recent architectures.

\begin{acks}
We would like to thank the Intel PSIRT for their timely and constructive feedback that helped us improve this paper, and the MSI PSIRT for recognizing our contribution in their Security Advisor Hall of Fame.
%    This document is derived from previous conferences, in particular MICRO 2013, ASPLOS 2015, MICRO 2015-2024, ISCA 2025, as well as SIGARCH/TCCA's Recommended Best Practices for the Conference Reviewing Process. 
\end{acks}

%%%%%%% -- PAPER CONTENT ENDS -- %%%%%%%%

%%
%% The next two lines define the bibliography style to be used, and
%% the bibliography file.
% \balance
\bibliographystyle{ACM-Reference-Format}
\bibliography{sample-base}

%%% -*-BibTeX-*-
%%% Do NOT edit. File created by BibTeX with style
%%% ACM-Reference-Format-Journals [18-Jan-2012].

\begin{thebibliography}{77}

%%% ====================================================================
%%% NOTE TO THE USER: you can override these defaults by providing
%%% customized versions of any of these macros before the \bibliography
%%% command.  Each of them MUST provide its own final punctuation,
%%% except for \shownote{}, \showDOI{}, and \showURL{}.  The latter two
%%% do not use final punctuation, in order to avoid confusing it with
%%% the Web address.
%%%
%%% To suppress output of a particular field, define its macro to expand
%%% to an empty string, or better, \unskip, like this:
%%%
%%% \newcommand{\showDOI}[1]{\unskip}   % LaTeX syntax
%%%
%%% \def \showDOI #1{\unskip}           % plain TeX syntax
%%%
%%% ====================================================================

\ifx \showCODEN    \undefined \def \showCODEN     #1{\unskip}     \fi
\ifx \showDOI      \undefined \def \showDOI       #1{#1}\fi
\ifx \showISBNx    \undefined \def \showISBNx     #1{\unskip}     \fi
\ifx \showISBNxiii \undefined \def \showISBNxiii  #1{\unskip}     \fi
\ifx \showISSN     \undefined \def \showISSN      #1{\unskip}     \fi
\ifx \showLCCN     \undefined \def \showLCCN      #1{\unskip}     \fi
\ifx \shownote     \undefined \def \shownote      #1{#1}          \fi
\ifx \showarticletitle \undefined \def \showarticletitle #1{#1}   \fi
\ifx \showURL      \undefined \def \showURL       {\relax}        \fi
% The following commands are used for tagged output and should be
% invisible to TeX
\providecommand\bibfield[2]{#2}
\providecommand\bibinfo[2]{#2}
\providecommand\natexlab[1]{#1}
\providecommand\showeprint[2][]{arXiv:#2}

\bibitem[Aweke et~al\mbox{.}(2016)]%
        {ANVIL}
\bibfield{author}{\bibinfo{person}{Zelalem~Birhanu Aweke}, \bibinfo{person}{Salessawi~Ferede Yitbarek}, \bibinfo{person}{Rui Qiao}, \bibinfo{person}{Reetuparna Das}, \bibinfo{person}{Matthew Hicks}, \bibinfo{person}{Yossi Oren}, {and} \bibinfo{person}{Todd~M. Austin}.} \bibinfo{year}{2016}\natexlab{}.
\newblock \showarticletitle{{ANVIL:} Software-Based Protection Against Next-Generation Rowhammer Attacks}. In \bibinfo{booktitle}{\emph{Proceedings of the Twenty-First International Conference on Architectural Support for Programming Languages and Operating Systems, {ASPLOS} 2016, Atlanta, GA, USA, April 2-6, 2016}}.
\newblock
\urldef\tempurl%
\url{https://doi.org/10.1145/2872362.2872390}
\showURL{%
\tempurl}


\bibitem[Bosman et~al\mbox{.}(2016)]%
        {Dedup_Est_Machina}
\bibfield{author}{\bibinfo{person}{Erik Bosman}, \bibinfo{person}{Kaveh Razavi}, \bibinfo{person}{Herbert Bos}, {and} \bibinfo{person}{Cristiano Giuffrida}.} \bibinfo{year}{2016}\natexlab{}.
\newblock \showarticletitle{Dedup Est Machina: Memory Deduplication as an Advanced Exploitation Vector}. In \bibinfo{booktitle}{\emph{{IEEE} Symposium on Security and Privacy, {SP} 2016, San Jose, CA, USA, May 22-26, 2016}}.
\newblock
\urldef\tempurl%
\url{https://doi.org/10.1109/SP.2016.63}
\showURL{%
\tempurl}


\bibitem[Brasser et~al\mbox{.}(2017)]%
        {Cannot_Touch_This}
\bibfield{author}{\bibinfo{person}{Ferdinand Brasser}, \bibinfo{person}{Lucas Davi}, \bibinfo{person}{David Gens}, \bibinfo{person}{Christopher Liebchen}, {and} \bibinfo{person}{Ahmad{-}Reza Sadeghi}.} \bibinfo{year}{2017}\natexlab{}.
\newblock \showarticletitle{CAn't Touch This: Software-only Mitigation against Rowhammer Attacks targeting Kernel Memory}. In \bibinfo{booktitle}{\emph{26th {USENIX} Security Symposium, {USENIX} Security 2017, Vancouver, BC, Canada, August 16-18, 2017}}.
\newblock
\urldef\tempurl%
\url{https://www.usenix.org/conference/usenixsecurity17/technical-sessions/presentation/brasser}
\showURL{%
\tempurl}


\bibitem[Bölcskei et~al\mbox{.}(2025)]%
        {rubicon}
\bibfield{author}{\bibinfo{person}{Matej Bölcskei}, \bibinfo{person}{Patrick Jattke}, \bibinfo{person}{Johannes Wikner}, {and} \bibinfo{person}{Kaveh Razavi}.} \bibinfo{year}{2025}\natexlab{}.
\newblock \showarticletitle{{Rubicon: Precise Microarchitectural Attacks with Page-Granular Massaging}}. In \bibinfo{booktitle}{\emph{{EuroS\&P}}}.
\newblock
\urldef\tempurl%
\url{Paper=https://comsec.ethz.ch/wp-content/files/rubicon_eurosp25.pdf}
\showURL{%
\tempurl}


\bibitem[Chen et~al\mbox{.}(2025)]%
        {hyperhammer}
\bibfield{author}{\bibinfo{person}{Wei Chen}, \bibinfo{person}{Zhi Zhang}, \bibinfo{person}{Xin Zhang}, \bibinfo{person}{Qingni Shen}, \bibinfo{person}{Yuval Yarom}, \bibinfo{person}{Daniel Genkin}, \bibinfo{person}{Chen Yan}, {and} \bibinfo{person}{Zhe Wang}.} \bibinfo{year}{2025}\natexlab{}.
\newblock \showarticletitle{HyperHammer: Breaking Free from KVM-Enforced Isolation}. In \bibinfo{booktitle}{\emph{Proceedings of the 30th ACM International Conference on Architectural Support for Programming Languages and Operating Systems, Volume 2}}. \bibinfo{pages}{545--559}.
\newblock


\bibitem[Cheng et~al\mbox{.}(2024)]%
        {evict_spec_time}
\bibfield{author}{\bibinfo{person}{Shing Hing~William Cheng}, \bibinfo{person}{Chitchanok Chuengsatiansup}, \bibinfo{person}{Daniel Genkin}, \bibinfo{person}{Dallas McNeil}, \bibinfo{person}{Toby Murray}, \bibinfo{person}{Yuval Yarom}, {and} \bibinfo{person}{Zhiyuan Zhang}.} \bibinfo{year}{2024}\natexlab{}.
\newblock \showarticletitle{Evict+Spec+Time: Exploiting Out-of-Order Execution to Improve Cache-Timing Attacks}.
\newblock \bibinfo{journal}{\emph{{IACR} Trans. Cryptogr. Hardw. Embed. Syst.}} \bibinfo{volume}{2024}, \bibinfo{number}{3} (\bibinfo{year}{2024}), \bibinfo{pages}{224--248}.
\newblock
\urldef\tempurl%
\url{https://doi.org/10.46586/TCHES.V2024.I3.224-248}
\showDOI{\tempurl}


\bibitem[Cohen et~al\mbox{.}(2022)]%
        {HammerScope}
\bibfield{author}{\bibinfo{person}{Yaakov Cohen}, \bibinfo{person}{Kevin~Sam Tharayil}, \bibinfo{person}{Arie Haenel}, \bibinfo{person}{Daniel Genkin}, \bibinfo{person}{Angelos~D. Keromytis}, \bibinfo{person}{Yossi Oren}, {and} \bibinfo{person}{Yuval Yarom}.} \bibinfo{year}{2022}\natexlab{}.
\newblock \showarticletitle{HammerScope: Observing {DRAM} Power Consumption Using Rowhammer}. In \bibinfo{booktitle}{\emph{Proceedings of the 2022 {ACM} {SIGSAC} Conference on Computer and Communications Security, {CCS} 2022, Los Angeles, CA, USA, November 7-11, 2022}}.
\newblock
\urldef\tempurl%
\url{https://doi.org/10.1145/3548606.3560688}
\showURL{%
\tempurl}


\bibitem[Cojocar et~al\mbox{.}(2020)]%
        {Are_We_Susceptible_to_Rowhammer}
\bibfield{author}{\bibinfo{person}{Lucian Cojocar}, \bibinfo{person}{Jeremie~S. Kim}, \bibinfo{person}{Minesh Patel}, \bibinfo{person}{Lillian Tsai}, \bibinfo{person}{Stefan Saroiu}, \bibinfo{person}{Alec Wolman}, {and} \bibinfo{person}{Onur Mutlu}.} \bibinfo{year}{2020}\natexlab{}.
\newblock \showarticletitle{Are We Susceptible to Rowhammer? An End-to-End Methodology for Cloud Providers}. In \bibinfo{booktitle}{\emph{2020 {IEEE} Symposium on Security and Privacy, {SP} 2020, San Francisco, CA, USA, May 18-21, 2020}}.
\newblock
\urldef\tempurl%
\url{https://doi.org/10.1109/SP40000.2020.00085}
\showURL{%
\tempurl}


\bibitem[Cojocar et~al\mbox{.}(2019)]%
        {ECC}
\bibfield{author}{\bibinfo{person}{Lucian Cojocar}, \bibinfo{person}{Kaveh Razavi}, \bibinfo{person}{Cristiano Giuffrida}, {and} \bibinfo{person}{Herbert Bos}.} \bibinfo{year}{2019}\natexlab{}.
\newblock \showarticletitle{Exploiting Correcting Codes: On the Effectiveness of {ECC} Memory Against Rowhammer Attacks}. In \bibinfo{booktitle}{\emph{2019 {IEEE} Symposium on Security and Privacy, {SP} 2019, San Francisco, CA, USA, May 19-23, 2019}}.
\newblock
\urldef\tempurl%
\url{https://doi.org/10.1109/SP.2019.00089}
\showURL{%
\tempurl}


\bibitem[de~Ridder et~al\mbox{.}(2021)]%
        {smash}
\bibfield{author}{\bibinfo{person}{Finn de Ridder}, \bibinfo{person}{Pietro Frigo}, \bibinfo{person}{Emanuele Vannacci}, \bibinfo{person}{Herbert Bos}, \bibinfo{person}{Cristiano Giuffrida}, {and} \bibinfo{person}{Kaveh Razavi}.} \bibinfo{year}{2021}\natexlab{}.
\newblock \showarticletitle{{SMASH:} Synchronized Many-sided Rowhammer Attacks from JavaScript}. In \bibinfo{booktitle}{\emph{30th {USENIX} Security Symposium, {USENIX} Security 2021, August 11-13, 2021}}.
\newblock
\urldef\tempurl%
\url{https://www.usenix.org/conference/usenixsecurity21/presentation/ridder}
\showURL{%
\tempurl}


\bibitem[Dio et~al\mbox{.}(2025)]%
        {halfspectre}
\bibfield{author}{\bibinfo{person}{Andrea~Di Dio}, \bibinfo{person}{Math{\'{e}} Hertogh}, {and} \bibinfo{person}{Cristiano Giuffrida}.} \bibinfo{year}{2025}\natexlab{}.
\newblock \showarticletitle{Half Spectre, Full Exploit: Hardening Rowhammer Attacks with Half-Spectre Gadgets}. In \bibinfo{booktitle}{\emph{{IEEE} Symposium on Security and Privacy, {SP} 2025, San Francisco, CA, USA, May 12-15, 2025}}.
\newblock
\urldef\tempurl%
\url{https://doi.org/10.1109/SP61157.2025.00207}
\showURL{%
\tempurl}


\bibitem[Fahr et~al\mbox{.}(2022)]%
        {When_Frodo_Flips}
\bibfield{author}{\bibinfo{person}{Michael Fahr}, \bibinfo{person}{Hunter Kippen}, \bibinfo{person}{Andrew Kwong}, \bibinfo{person}{Thinh Dang}, \bibinfo{person}{Jacob Lichtinger}, \bibinfo{person}{Dana Dachman{-}Soled}, \bibinfo{person}{Daniel Genkin}, \bibinfo{person}{Alexander Nelson}, \bibinfo{person}{Ray~A. Perlner}, \bibinfo{person}{Arkady Yerukhimovich}, {and} \bibinfo{person}{Daniel Apon}.} \bibinfo{year}{2022}\natexlab{}.
\newblock \showarticletitle{When Frodo Flips: End-to-End Key Recovery on FrodoKEM via Rowhammer}. In \bibinfo{booktitle}{\emph{Proceedings of the 2022 {ACM} {SIGSAC} Conference on Computer and Communications Security, {CCS} 2022, Los Angeles, CA, USA, November 7-11, 2022}}.
\newblock
\urldef\tempurl%
\url{https://doi.org/10.1145/3548606.3560673}
\showURL{%
\tempurl}


\bibitem[Frigo et~al\mbox{.}(2018)]%
        {Grand_unit}
\bibfield{author}{\bibinfo{person}{Pietro Frigo}, \bibinfo{person}{Cristiano Giuffrida}, \bibinfo{person}{Herbert Bos}, {and} \bibinfo{person}{Kaveh Razavi}.} \bibinfo{year}{2018}\natexlab{}.
\newblock \showarticletitle{Grand Pwning Unit: Accelerating Microarchitectural Attacks with the {GPU}}. In \bibinfo{booktitle}{\emph{2018 {IEEE} Symposium on Security and Privacy, {SP} 2018, Proceedings, 21-23 May 2018, San Francisco, California, {USA}}}.
\newblock
\urldef\tempurl%
\url{https://doi.org/10.1109/SP.2018.00022}
\showURL{%
\tempurl}


\bibitem[Frigo et~al\mbox{.}(2020)]%
        {trrespass}
\bibfield{author}{\bibinfo{person}{Pietro Frigo}, \bibinfo{person}{Emanuele Vannacci}, \bibinfo{person}{Hasan Hassan}, \bibinfo{person}{Victor van~der Veen}, \bibinfo{person}{Onur Mutlu}, \bibinfo{person}{Cristiano Giuffrida}, \bibinfo{person}{Herbert Bos}, {and} \bibinfo{person}{Kaveh Razavi}.} \bibinfo{year}{2020}\natexlab{}.
\newblock \showarticletitle{TRRespass: Exploiting the Many Sides of Target Row Refresh}. In \bibinfo{booktitle}{\emph{2020 {IEEE} Symposium on Security and Privacy, {SP} 2020, San Francisco, CA, USA, May 18-21, 2020}}.
\newblock
\urldef\tempurl%
\url{https://doi.org/10.1109/SP40000.2020.00090}
\showURL{%
\tempurl}


\bibitem[Gruss et~al\mbox{.}(2018)]%
        {Another_Flip_in_the_Wall_of_Rowhammer_Defenses}
\bibfield{author}{\bibinfo{person}{Daniel Gruss}, \bibinfo{person}{Moritz Lipp}, \bibinfo{person}{Michael Schwarz}, \bibinfo{person}{Daniel Genkin}, \bibinfo{person}{Jonas Juffinger}, \bibinfo{person}{Sioli O'Connell}, \bibinfo{person}{Wolfgang Schoechl}, {and} \bibinfo{person}{Yuval Yarom}.} \bibinfo{year}{2018}\natexlab{}.
\newblock \showarticletitle{Another Flip in the Wall of Rowhammer Defenses}. In \bibinfo{booktitle}{\emph{2018 {IEEE} Symposium on Security and Privacy, {SP} 2018, Proceedings, 21-23 May 2018, San Francisco, California, {USA}}}.
\newblock
\urldef\tempurl%
\url{https://doi.org/10.1109/SP.2018.00031}
\showURL{%
\tempurl}


\bibitem[Gruss et~al\mbox{.}(2016)]%
        {Rowhammerjs}
\bibfield{author}{\bibinfo{person}{Daniel Gruss}, \bibinfo{person}{Cl{\'{e}}mentine Maurice}, {and} \bibinfo{person}{Stefan Mangard}.} \bibinfo{year}{2016}\natexlab{}.
\newblock \showarticletitle{Rowhammer.js: {A} Remote Software-Induced Fault Attack in JavaScript}. In \bibinfo{booktitle}{\emph{Detection of Intrusions and Malware, and Vulnerability Assessment - 13th International Conference, {DIMVA} 2016, San Sebasti{\'{a}}n, Spain, July 7-8, 2016, Proceedings}}.
\newblock
\urldef\tempurl%
\url{https://doi.org/10.1007/978-3-319-40667-1\_15}
\showURL{%
\tempurl}


\bibitem[Guo et~al\mbox{.}(2022)]%
        {adversarial_prefetch}
\bibfield{author}{\bibinfo{person}{Yanan Guo}, \bibinfo{person}{Andrew Zigerelli}, \bibinfo{person}{Youtao Zhang}, {and} \bibinfo{person}{Jun Yang}.} \bibinfo{year}{2022}\natexlab{}.
\newblock \showarticletitle{Adversarial Prefetch: New Cross-Core Cache Side Channel Attacks}. In \bibinfo{booktitle}{\emph{43rd {IEEE} Symposium on Security and Privacy, {SP} 2022, San Francisco, CA, USA, May 22-26, 2022}}.
\newblock
\urldef\tempurl%
\url{https://doi.org/10.1109/SP46214.2022.9833692}
\showURL{%
\tempurl}


\bibitem[Hassan et~al\mbox{.}(2021)]%
        {utrr}
\bibfield{author}{\bibinfo{person}{Hasan Hassan}, \bibinfo{person}{Yahya~Can Tugrul}, \bibinfo{person}{Jeremie~S. Kim}, \bibinfo{person}{Victor van~der Veen}, \bibinfo{person}{Kaveh Razavi}, {and} \bibinfo{person}{Onur Mutlu}.} \bibinfo{year}{2021}\natexlab{}.
\newblock \showarticletitle{Uncovering In-DRAM RowHammer Protection Mechanisms: {A} New Methodology, Custom RowHammer Patterns, and Implications}. In \bibinfo{booktitle}{\emph{{MICRO} '21: 54th Annual {IEEE/ACM} International Symposium on Microarchitecture, Virtual Event, Greece, October 18-22, 2021}}.
\newblock
\urldef\tempurl%
\url{https://doi.org/10.1145/3466752.3480110}
\showURL{%
\tempurl}


\bibitem[He et~al\mbox{.}(2025)]%
        {WhistleBlower}
\bibfield{author}{\bibinfo{person}{Wei He}, \bibinfo{person}{Zhi Zhang}, \bibinfo{person}{Yueqiang Cheng}, \bibinfo{person}{Wenhao Wang}, \bibinfo{person}{Wei Song}, \bibinfo{person}{Yansong Gao}, \bibinfo{person}{Qifei Zhang}, \bibinfo{person}{Kang Li}, \bibinfo{person}{Dongxi Liu}, {and} \bibinfo{person}{Surya Nepal}.} \bibinfo{year}{2025}\natexlab{}.
\newblock \showarticletitle{WhistleBlower: {A} System-Level Empirical Study on RowHammer}.
\newblock \bibinfo{journal}{\emph{{IEEE} Trans. Computers}} \bibinfo{volume}{74}, \bibinfo{number}{3} (\bibinfo{year}{2025}), \bibinfo{pages}{805--819}.
\newblock
\urldef\tempurl%
\url{https://doi.org/10.1109/TC.2023.3235973}
\showDOI{\tempurl}


\bibitem[Helm et~al\mbox{.}(2020)]%
        {Reliable_Reverse_Engineering_of_Intel_DRAM_Addressing_Using_Performance_Counters}
\bibfield{author}{\bibinfo{person}{Christian Helm}, \bibinfo{person}{Soramichi Akiyama}, {and} \bibinfo{person}{Kenjiro Taura}.} \bibinfo{year}{2020}\natexlab{}.
\newblock \showarticletitle{Reliable Reverse Engineering of Intel {DRAM} Addressing Using Performance Counters}. In \bibinfo{booktitle}{\emph{28th International Symposium on Modeling, Analysis, and Simulation of Computer and Telecommunication Systems, {MASCOTS} 2020, Nice, France, November 17-19, 2020}}.
\newblock
\urldef\tempurl%
\url{https://doi.org/10.1109/MASCOTS50786.2020.9285962}
\showURL{%
\tempurl}


\bibitem[Hof and Nieh(2022)]%
        {osdi_blackbox}
\bibfield{author}{\bibinfo{person}{Alexander~Van't Hof} {and} \bibinfo{person}{Jason Nieh}.} \bibinfo{year}{2022}\natexlab{}.
\newblock \showarticletitle{BlackBox: {A} Container Security Monitor for Protecting Containers on Untrusted Operating Systems}. In \bibinfo{booktitle}{\emph{16th {USENIX} Symposium on Operating Systems Design and Implementation, {OSDI} 2022, Carlsbad, CA, USA, July 11-13, 2022}}.
\newblock
\urldef\tempurl%
\url{https://www.usenix.org/conference/osdi22/presentation/vant-hof}
\showURL{%
\tempurl}


\bibitem[Hong et~al\mbox{.}(2019)]%
        {terminal_brain}
\bibfield{author}{\bibinfo{person}{Sanghyun Hong}, \bibinfo{person}{Pietro Frigo}, \bibinfo{person}{Yigitcan Kaya}, \bibinfo{person}{Cristiano Giuffrida}, {and} \bibinfo{person}{Tudor Dumitras}.} \bibinfo{year}{2019}\natexlab{}.
\newblock \showarticletitle{Terminal Brain Damage: Exposing the Graceless Degradation in Deep Neural Networks Under Hardware Fault Attacks}. In \bibinfo{booktitle}{\emph{28th {USENIX} Security Symposium, {USENIX} Security 2019, Santa Clara, CA, USA, August 14-16, 2019}}.
\newblock
\urldef\tempurl%
\url{https://www.usenix.org/conference/usenixsecurity19/presentation/hong}
\showURL{%
\tempurl}


\bibitem[Hunt et~al\mbox{.}(2016)]%
        {osdi_ryoan}
\bibfield{author}{\bibinfo{person}{Tyler Hunt}, \bibinfo{person}{Zhiting Zhu}, \bibinfo{person}{Yuanzhong Xu}, \bibinfo{person}{Simon Peter}, {and} \bibinfo{person}{Emmett Witchel}.} \bibinfo{year}{2016}\natexlab{}.
\newblock \showarticletitle{Ryoan: {A} Distributed Sandbox for Untrusted Computation on Secret Data}. In \bibinfo{booktitle}{\emph{12th {USENIX} Symposium on Operating Systems Design and Implementation, {OSDI} 2016, Savannah, GA, USA, November 2-4, 2016}}.
\newblock
\urldef\tempurl%
\url{https://www.usenix.org/conference/osdi16/technical-sessions/presentation/hunt}
\showURL{%
\tempurl}


\bibitem[Intel(2023)]%
        {12gen_manual}
\bibfield{author}{\bibinfo{person}{Intel}.} \bibinfo{year}{2023}\natexlab{}.
\newblock \bibinfo{title}{12th Generation Intel{\textregistered} Core™ Processors}.
\newblock
\newblock
\urldef\tempurl%
\url{https://cdrdv2-public.intel.com/682436/682436-031.pdf}
\showURL{%
\tempurl}


\bibitem[Intel(2024a)]%
        {intel_arch_opt}
\bibfield{author}{\bibinfo{person}{Intel}.} \bibinfo{year}{2024}\natexlab{a}.
\newblock \bibinfo{title}{Intel{\textregistered} 64 and IA-32 Architectures Optimization Reference Manual}.
\newblock
\newblock
\urldef\tempurl%
\url{https://cdrdv2.intel.com/v1/dl/getContent/671488}
\showURL{%
\tempurl}


\bibitem[Intel(2024b)]%
        {intel_soft_dev}
\bibfield{author}{\bibinfo{person}{Intel}.} \bibinfo{year}{2024}\natexlab{b}.
\newblock \bibinfo{title}{Intel{\textregistered} 64 and IA-32 Architectures Software Developer’s Manual}.
\newblock
\newblock
\urldef\tempurl%
\url{https://cdrdv2-public.intel.com/825743/325462-sdm-vol-1-2abcd-3abcd-4.pdf}
\showURL{%
\tempurl}


\bibitem[Islam et~al\mbox{.}(2019)]%
        {SPOILER}
\bibfield{author}{\bibinfo{person}{Saad Islam}, \bibinfo{person}{Ahmad Moghimi}, \bibinfo{person}{Ida Bruhns}, \bibinfo{person}{Moritz Krebbel}, \bibinfo{person}{Berk G{\"{u}}lmezoglu}, \bibinfo{person}{Thomas Eisenbarth}, {and} \bibinfo{person}{Berk Sunar}.} \bibinfo{year}{2019}\natexlab{}.
\newblock \showarticletitle{{SPOILER:} Speculative Load Hazards Boost Rowhammer and Cache Attacks}. In \bibinfo{booktitle}{\emph{28th {USENIX} Security Symposium, {USENIX} Security 2019, Santa Clara, CA, USA, August 14-16, 2019}}.
\newblock
\urldef\tempurl%
\url{https://www.usenix.org/conference/usenixsecurity19/presentation/islam}
\showURL{%
\tempurl}


\bibitem[Jang et~al\mbox{.}(2017)]%
        {SGX-Bomb}
\bibfield{author}{\bibinfo{person}{Yeongjin Jang}, \bibinfo{person}{Jae{-}Hyuk Lee}, \bibinfo{person}{Sangho Lee}, {and} \bibinfo{person}{Taesoo Kim}.} \bibinfo{year}{2017}\natexlab{}.
\newblock \showarticletitle{SGX-Bomb: Locking Down the Processor via Rowhammer Attack}. In \bibinfo{booktitle}{\emph{Proceedings of the 2nd Workshop on System Software for Trusted Execution, SysTEX@SOSP 2017, Shanghai, China, October 28, 2017}}.
\newblock
\urldef\tempurl%
\url{https://doi.org/10.1145/3152701.3152709}
\showURL{%
\tempurl}


\bibitem[Jattke et~al\mbox{.}(2022)]%
        {blacksmith}
\bibfield{author}{\bibinfo{person}{Patrick Jattke}, \bibinfo{person}{Victor van~der Veen}, \bibinfo{person}{Pietro Frigo}, \bibinfo{person}{Stijn Gunter}, {and} \bibinfo{person}{Kaveh Razavi}.} \bibinfo{year}{2022}\natexlab{}.
\newblock \showarticletitle{{BLACKSMITH:} Scalable Rowhammering in the Frequency Domain}. In \bibinfo{booktitle}{\emph{43rd {IEEE} Symposium on Security and Privacy, {SP} 2022, San Francisco, CA, USA, May 22-26, 2022}}.
\newblock
\urldef\tempurl%
\url{https://doi.org/10.1109/SP46214.2022.9833772}
\showURL{%
\tempurl}


\bibitem[Jattke et~al\mbox{.}(2024)]%
        {zenhammer}
\bibfield{author}{\bibinfo{person}{Patrick Jattke}, \bibinfo{person}{Max Wipfli}, \bibinfo{person}{Flavien Solt}, \bibinfo{person}{Michele Marazzi}, \bibinfo{person}{Matej B{\"{o}}lcskei}, {and} \bibinfo{person}{Kaveh Razavi}.} \bibinfo{year}{2024}\natexlab{}.
\newblock \showarticletitle{ZenHammer: Rowhammer Attacks on {AMD} Zen-based Platforms}. In \bibinfo{booktitle}{\emph{33rd {USENIX} Security Symposium, {USENIX} Security 2024, Philadelphia, PA, USA, August 14-16, 2024}}.
\newblock
\urldef\tempurl%
\url{https://www.usenix.org/conference/usenixsecurity24/presentation/jattke}
\showURL{%
\tempurl}


\bibitem[{JEDEC}(2020)]%
        {jedec_ddr5}
\bibfield{author}{\bibinfo{person}{{JEDEC}}.} \bibinfo{year}{2020}\natexlab{}.
\newblock \bibinfo{title}{DDR5 SDRAM Specification}.
\newblock
\newblock
\urldef\tempurl%
\url{https://www.jedec.org/standards-documents/docs/jesd79-5c01}
\showURL{%
\tempurl}


\bibitem[Kaczmarski(2014)]%
        {intel_xeon}
\bibfield{author}{\bibinfo{person}{Marcin Kaczmarski}.} \bibinfo{year}{2014}\natexlab{}.
\newblock \showarticletitle{Thoughts on intel xeon e5-2600 v2 product family performance optimisation--component selection guidelines}.
\newblock \bibinfo{journal}{\emph{Santa Clara, CA, USA: Intel}} (\bibinfo{year}{2014}).
\newblock


\bibitem[Kang et~al\mbox{.}(2024)]%
        {sledgehammer}
\bibfield{author}{\bibinfo{person}{Ingab Kang}, \bibinfo{person}{Walter Wang}, \bibinfo{person}{Jason Kim}, \bibinfo{person}{Stephan van Schaik}, \bibinfo{person}{Youssef Tobah}, \bibinfo{person}{Daniel Genkin}, \bibinfo{person}{Andrew Kwong}, {and} \bibinfo{person}{Yuval Yarom}.} \bibinfo{year}{2024}\natexlab{}.
\newblock \showarticletitle{SledgeHammer: Amplifying Rowhammer via Bank-level Parallelism}. In \bibinfo{booktitle}{\emph{33rd {USENIX} Security Symposium, {USENIX} Security 2024, Philadelphia, PA, USA, August 14-16, 2024}}.
\newblock
\urldef\tempurl%
\url{https://www.usenix.org/conference/usenixsecurity24/presentation/kang}
\showURL{%
\tempurl}


\bibitem[Kim et~al\mbox{.}(2023)]%
        {cube}
\bibfield{author}{\bibinfo{person}{Michael~Jaemin Kim}, \bibinfo{person}{Minbok Wi}, \bibinfo{person}{Jaehyun Park}, \bibinfo{person}{Seoyoung Ko}, \bibinfo{person}{Jaeyoung Choi}, \bibinfo{person}{Hwayong Nam}, \bibinfo{person}{Nam~Sung Kim}, \bibinfo{person}{Jung~Ho Ahn}, {and} \bibinfo{person}{Eojin Lee}.} \bibinfo{year}{2023}\natexlab{}.
\newblock \showarticletitle{How to Kill the Second Bird with One {ECC:} The Pursuit of Row Hammer Resilient {DRAM}}. In \bibinfo{booktitle}{\emph{Proceedings of the 56th Annual {IEEE/ACM} International Symposium on Microarchitecture, {MICRO} 2023, Toronto, ON, Canada, 28 October 2023 - 1 November 2023}}.
\newblock
\urldef\tempurl%
\url{https://doi.org/10.1145/3613424.3623777}
\showURL{%
\tempurl}


\bibitem[Kim et~al\mbox{.}(2014)]%
        {Fliping}
\bibfield{author}{\bibinfo{person}{Yoongu Kim}, \bibinfo{person}{Ross Daly}, \bibinfo{person}{Jeremie~S. Kim}, \bibinfo{person}{Chris Fallin}, \bibinfo{person}{Ji{-}Hye Lee}, \bibinfo{person}{Donghyuk Lee}, \bibinfo{person}{Chris Wilkerson}, \bibinfo{person}{Konrad Lai}, {and} \bibinfo{person}{Onur Mutlu}.} \bibinfo{year}{2014}\natexlab{}.
\newblock \showarticletitle{Flipping bits in memory without accessing them: An experimental study of {DRAM} disturbance errors}. In \bibinfo{booktitle}{\emph{{ACM/IEEE} 41st International Symposium on Computer Architecture, {ISCA} 2014, Minneapolis, MN, USA, June 14-18, 2014}}.
\newblock
\urldef\tempurl%
\url{https://doi.org/10.1109/ISCA.2014.6853210}
\showURL{%
\tempurl}


\bibitem[Kobalicek(2023)]%
        {asmjit}
\bibfield{author}{\bibinfo{person}{Petr Kobalicek}.} \bibinfo{year}{2023}\natexlab{}.
\newblock \bibinfo{title}{asmjit: Low-latency machine code generation}.
\newblock
\newblock
\urldef\tempurl%
\url{https://asmjit.com/}
\showURL{%
\tempurl}


\bibitem[Kogler et~al\mbox{.}(2022)]%
        {Half-Double}
\bibfield{author}{\bibinfo{person}{Andreas Kogler}, \bibinfo{person}{Jonas Juffinger}, \bibinfo{person}{Salman Qazi}, \bibinfo{person}{Yoongu Kim}, \bibinfo{person}{Moritz Lipp}, \bibinfo{person}{Nicolas Boichat}, \bibinfo{person}{Eric Shiu}, \bibinfo{person}{Mattias Nissler}, {and} \bibinfo{person}{Daniel Gruss}.} \bibinfo{year}{2022}\natexlab{}.
\newblock \showarticletitle{Half-Double: Hammering From the Next Row Over}. In \bibinfo{booktitle}{\emph{31st {USENIX} Security Symposium, {USENIX} Security 2022, Boston, MA, USA, August 10-12, 2022}}.
\newblock
\urldef\tempurl%
\url{https://www.usenix.org/conference/usenixsecurity22/presentation/kogler-half-double}
\showURL{%
\tempurl}


\bibitem[Konoth et~al\mbox{.}(2018)]%
        {ZebRAM}
\bibfield{author}{\bibinfo{person}{Radhesh~Krishnan Konoth}, \bibinfo{person}{Marco Oliverio}, \bibinfo{person}{Andrei Tatar}, \bibinfo{person}{Dennis Andriesse}, \bibinfo{person}{Herbert Bos}, \bibinfo{person}{Cristiano Giuffrida}, {and} \bibinfo{person}{Kaveh Razavi}.} \bibinfo{year}{2018}\natexlab{}.
\newblock \showarticletitle{ZebRAM: Comprehensive and Compatible Software Protection Against Rowhammer Attacks}. In \bibinfo{booktitle}{\emph{13th {USENIX} Symposium on Operating Systems Design and Implementation, {OSDI} 2018, Carlsbad, CA, USA, October 8-10, 2018}}.
\newblock
\urldef\tempurl%
\url{https://www.usenix.org/conference/osdi18/presentation/konoth}
\showURL{%
\tempurl}


\bibitem[K{\"{u}}hn et~al\mbox{.}(2024)]%
        {How_to_Be_Fast_and_Not_Furious}
\bibfield{author}{\bibinfo{person}{Roland K{\"{u}}hn}, \bibinfo{person}{Jan M{\"{u}}hlig}, {and} \bibinfo{person}{Jens Teubner}.} \bibinfo{year}{2024}\natexlab{}.
\newblock \showarticletitle{How to Be Fast and Not Furious: Looking Under the Hood of {CPU} Cache Prefetching}. In \bibinfo{booktitle}{\emph{Proceedings of the 20th International Workshop on Data Management on New Hardware, DaMoN 2024, Santiago, Chile, 10 June 2024}}.
\newblock
\urldef\tempurl%
\url{https://doi.org/10.1145/3662010.3663451}
\showURL{%
\tempurl}


\bibitem[Kwong et~al\mbox{.}(2020)]%
        {rambleed}
\bibfield{author}{\bibinfo{person}{Andrew Kwong}, \bibinfo{person}{Daniel Genkin}, \bibinfo{person}{Daniel Gruss}, {and} \bibinfo{person}{Yuval Yarom}.} \bibinfo{year}{2020}\natexlab{}.
\newblock \showarticletitle{RAMBleed: Reading Bits in Memory Without Accessing Them}. In \bibinfo{booktitle}{\emph{2020 {IEEE} Symposium on Security and Privacy, {SP} 2020, San Francisco, CA, USA, May 18-21, 2020}}.
\newblock
\urldef\tempurl%
\url{https://doi.org/10.1109/SP40000.2020.00020}
\showURL{%
\tempurl}


\bibitem[Lee et~al\mbox{.}(2019)]%
        {TWiCe}
\bibfield{author}{\bibinfo{person}{Eojin Lee}, \bibinfo{person}{Ingab Kang}, \bibinfo{person}{Sukhan Lee}, \bibinfo{person}{G.~Edward Suh}, {and} \bibinfo{person}{Jung~Ho Ahn}.} \bibinfo{year}{2019}\natexlab{}.
\newblock \showarticletitle{TWiCe: preventing row-hammering by exploiting time window counters}. In \bibinfo{booktitle}{\emph{Proceedings of the 46th International Symposium on Computer Architecture, {ISCA} 2019, Phoenix, AZ, USA, June 22-26, 2019}}.
\newblock
\urldef\tempurl%
\url{https://doi.org/10.1145/3307650.3322232}
\showURL{%
\tempurl}


\bibitem[Lee(2014)]%
        {Samsung}
\bibfield{author}{\bibinfo{person}{J.-B. Lee}.} \bibinfo{year}{2014}\natexlab{}.
\newblock \bibinfo{title}{“Green Memory Solution,” in Samsung Electronics, Investor’s Forum, 2014}.
\newblock
\newblock


\bibitem[Li et~al\mbox{.}(2024)]%
        {Yes_One-Bit-Flip_Matters!}
\bibfield{author}{\bibinfo{person}{Shaofeng Li}, \bibinfo{person}{Xinyu Wang}, \bibinfo{person}{Minhui Xue}, \bibinfo{person}{Haojin Zhu}, \bibinfo{person}{Zhi Zhang}, \bibinfo{person}{Yansong Gao}, \bibinfo{person}{Wen Wu}, {and} \bibinfo{person}{Xuemin~(Sherman) Shen}.} \bibinfo{year}{2024}\natexlab{}.
\newblock \showarticletitle{Yes, One-Bit-Flip Matters! Universal {DNN} Model Inference Depletion with Runtime Code Fault Injection}. In \bibinfo{booktitle}{\emph{33rd {USENIX} Security Symposium, {USENIX} Security 2024, Philadelphia, PA, USA, August 14-16, 2024}}.
\newblock
\urldef\tempurl%
\url{https://www.usenix.org/conference/usenixsecurity24/presentation/li-shaofeng}
\showURL{%
\tempurl}


\bibitem[Lipp et~al\mbox{.}(2022)]%
        {amd_prefetch}
\bibfield{author}{\bibinfo{person}{Moritz Lipp}, \bibinfo{person}{Daniel Gruss}, {and} \bibinfo{person}{Michael Schwarz}.} \bibinfo{year}{2022}\natexlab{}.
\newblock \showarticletitle{{AMD} Prefetch Attacks through Power and Time}. In \bibinfo{booktitle}{\emph{31st {USENIX} Security Symposium, {USENIX} Security 2022, Boston, MA, USA, August 10-12, 2022}}.
\newblock
\urldef\tempurl%
\url{https://www.usenix.org/conference/usenixsecurity22/presentation/lipp}
\showURL{%
\tempurl}


\bibitem[Luo et~al\mbox{.}(2023)]%
        {RowPress}
\bibfield{author}{\bibinfo{person}{Haocong Luo}, \bibinfo{person}{Ataberk Olgun}, \bibinfo{person}{Abdullah~Giray Yaglik{\c{c}}i}, \bibinfo{person}{Yahya~Can Tugrul}, \bibinfo{person}{Steve Rhyner}, \bibinfo{person}{Meryem~Banu Cavlak}, \bibinfo{person}{Jo{\"{e}}l Lindegger}, \bibinfo{person}{Mohammad Sadrosadati}, {and} \bibinfo{person}{Onur Mutlu}.} \bibinfo{year}{2023}\natexlab{}.
\newblock \showarticletitle{RowPress: Amplifying Read Disturbance in Modern {DRAM} Chips}. In \bibinfo{booktitle}{\emph{Proceedings of the 50th Annual International Symposium on Computer Architecture, {ISCA} 2023, Orlando, FL, USA, June 17-21, 2023}}.
\newblock
\urldef\tempurl%
\url{https://doi.org/10.1145/3579371.3589063}
\showURL{%
\tempurl}


\bibitem[Marazzi and Razavi(2024)]%
        {risc-h}
\bibfield{author}{\bibinfo{person}{Michele Marazzi} {and} \bibinfo{person}{Kaveh Razavi}.} \bibinfo{year}{2024}\natexlab{}.
\newblock \showarticletitle{{RISC-H}: Rowhammer Attacks on {RISC-V}}. In \bibinfo{booktitle}{\emph{4th Workshop on DRAM Security (DRAMSec), 2024}}.
\newblock
\urldef\tempurl%
\url{https://www.research-collection.ethz.ch/handle/20.500.11850/678065}
\showURL{%
\tempurl}


\bibitem[McCalpin(2006)]%
        {stream}
\bibfield{author}{\bibinfo{person}{John McCalpin}.} \bibinfo{year}{2006}\natexlab{}.
\newblock \showarticletitle{STREAM: Sustainable memory bandwidth in high performance computers}.
\newblock  (\bibinfo{year}{2006}).
\newblock
\urldef\tempurl%
\url{http://www.cs.virginia.edu/stream/}
\showURL{%
\tempurl}


\bibitem[Micron(2016)]%
        {Micron}
\bibfield{author}{\bibinfo{person}{Micron}.} \bibinfo{year}{2016}\natexlab{}.
\newblock \bibinfo{title}{“DDR4 SDRAM Datasheet,” p. 380, 2016.}
\newblock
\newblock


\bibitem[Micron(2020)]%
        {Micron_inc}
\bibfield{author}{\bibinfo{person}{Inc Micron}.} \bibinfo{year}{2020}\natexlab{}.
\newblock \bibinfo{title}{8Gb: x4, x8, x16 DDR4 SDRAM Features-Excessive Row Activation}.
\newblock
\newblock
\urldef\tempurl%
\url{https://www.micron.com/products/dram/ddr4-sdram}
\showURL{%
\tempurl}


\bibitem[Nam et~al\mbox{.}(2024)]%
        {DRAMScope}
\bibfield{author}{\bibinfo{person}{Hwayong Nam}, \bibinfo{person}{Seungmin Baek}, \bibinfo{person}{Minbok Wi}, \bibinfo{person}{Michael~Jaemin Kim}, \bibinfo{person}{Jaehyun Park}, \bibinfo{person}{Chihun Song}, \bibinfo{person}{Nam~Sung Kim}, {and} \bibinfo{person}{Jung~Ho Ahn}.} \bibinfo{year}{2024}\natexlab{}.
\newblock \showarticletitle{DRAMScope: Uncovering {DRAM} Microarchitecture and Characteristics by Issuing Memory Commands}. In \bibinfo{booktitle}{\emph{51st {ACM/IEEE} Annual International Symposium on Computer Architecture, {ISCA} 2024, Buenos Aires, Argentina, June 29 - July 3, 2024}}.
\newblock
\urldef\tempurl%
\url{https://doi.org/10.1109/ISCA59077.2024.00083}
\showURL{%
\tempurl}


\bibitem[Nazaraliyev et~al\mbox{.}(2025)]%
        {not_so_refreshing}
\bibfield{author}{\bibinfo{person}{Ravan Nazaraliyev}, \bibinfo{person}{Yicheng Zhang}, \bibinfo{person}{Sankha~Baran Dutta}, \bibinfo{person}{Andres Marquez}, \bibinfo{person}{Kevin Barker}, {and} \bibinfo{person}{Nael Abu-Ghazaleh}.} \bibinfo{year}{2025}\natexlab{}.
\newblock \showarticletitle{Not so Refreshing: Attacking {GPUs} using RFM Rowhammer Mitigation}. In \bibinfo{booktitle}{\emph{34th {USENIX} Security Symposium, {USENIX} Security 2025, Seattle, WA, USA, August 13-15, 2025}}.
\newblock
\urldef\tempurl%
\url{https://www.usenix.org/conference/usenixsecurity25/presentation/nazaraliyev}
\showURL{%
\tempurl}


\bibitem[Orosa et~al\mbox{.}(2021)]%
        {deeper_look_into}
\bibfield{author}{\bibinfo{person}{Lois Orosa}, \bibinfo{person}{Abdullah~Giray Yaglik{\c{c}}i}, \bibinfo{person}{Haocong Luo}, \bibinfo{person}{Ataberk Olgun}, \bibinfo{person}{Jisung Park}, \bibinfo{person}{Hasan Hassan}, \bibinfo{person}{Minesh Patel}, \bibinfo{person}{Jeremie~S. Kim}, {and} \bibinfo{person}{Onur Mutlu}.} \bibinfo{year}{2021}\natexlab{}.
\newblock \showarticletitle{A Deeper Look into RowHammer's Sensitivities: Experimental Analysis of Real {DRAM} Chipsand Implications on Future Attacks and Defenses}. In \bibinfo{booktitle}{\emph{{MICRO} '21: 54th Annual {IEEE/ACM} International Symposium on Microarchitecture, Virtual Event, Greece, October 18-22, 2021}}.
\newblock
\urldef\tempurl%
\url{https://doi.org/10.1145/3466752.3480069}
\showURL{%
\tempurl}


\bibitem[Park et~al\mbox{.}(2024)]%
        {taco_trusted_serverless}
\bibfield{author}{\bibinfo{person}{Joongun Park}, \bibinfo{person}{Seunghyo Kang}, \bibinfo{person}{Sanghyeon Lee}, \bibinfo{person}{Taehoon Kim}, \bibinfo{person}{Jongse Park}, \bibinfo{person}{Youngjin Kwon}, {and} \bibinfo{person}{Jaehyuk Huh}.} \bibinfo{year}{2024}\natexlab{}.
\newblock \showarticletitle{Hardware-hardened Sandbox Enclaves for Trusted Serverless Computing}.
\newblock \bibinfo{journal}{\emph{{ACM} Trans. Archit. Code Optim.}} \bibinfo{volume}{21}, \bibinfo{number}{1} (\bibinfo{year}{2024}), \bibinfo{pages}{13:1--13:25}.
\newblock
\urldef\tempurl%
\url{https://doi.org/10.1145/3632954}
\showDOI{\tempurl}


\bibitem[Park et~al\mbox{.}(2020)]%
        {Graphene}
\bibfield{author}{\bibinfo{person}{Yeonhong Park}, \bibinfo{person}{Woosuk Kwon}, \bibinfo{person}{Eojin Lee}, \bibinfo{person}{Tae~Jun Ham}, \bibinfo{person}{Jung~Ho Ahn}, {and} \bibinfo{person}{Jae~W. Lee}.} \bibinfo{year}{2020}\natexlab{}.
\newblock \showarticletitle{Graphene: Strong yet Lightweight Row Hammer Protection}. In \bibinfo{booktitle}{\emph{53rd Annual {IEEE/ACM} International Symposium on Microarchitecture, {MICRO} 2020, Athens, Greece, October 17-21, 2020}}.
\newblock
\urldef\tempurl%
\url{https://doi.org/10.1109/MICRO50266.2020.00014}
\showURL{%
\tempurl}


\bibitem[Peng et~al\mbox{.}(2025)]%
        {atc_asterinas}
\bibfield{author}{\bibinfo{person}{Yuke Peng}, \bibinfo{person}{Hongliang Tian}, \bibinfo{person}{Junyang Zhang}, \bibinfo{person}{Ruihan Li}, \bibinfo{person}{Chengjun Chen}, \bibinfo{person}{Jianfeng Jiang}, \bibinfo{person}{Jinyi Xian}, \bibinfo{person}{Xiaolin Wang}, \bibinfo{person}{Chenren Xu}, \bibinfo{person}{Diyu Zhou}, \bibinfo{person}{Yingwei Luo}, \bibinfo{person}{Shoumeng Yan}, {and} \bibinfo{person}{Yinqian Zhang}.} \bibinfo{year}{2025}\natexlab{}.
\newblock \showarticletitle{{ASTERINAS:} {A} Linux ABI-Compatible, Rust-Based Framekernel {OS} with a Small and Sound {TCB}}. In \bibinfo{booktitle}{\emph{Proceedings of the 2025 {USENIX} Annual Technical Conference, {USENIX} {ATC} 2025, Boston, MA, USA, July 7-9, 2025}}.
\newblock
\urldef\tempurl%
\url{https://www.usenix.org/conference/atc25/presentation/peng-yuke}
\showURL{%
\tempurl}


\bibitem[Pessl et~al\mbox{.}(2016)]%
        {drama}
\bibfield{author}{\bibinfo{person}{Peter Pessl}, \bibinfo{person}{Daniel Gruss}, \bibinfo{person}{Cl{\'{e}}mentine Maurice}, \bibinfo{person}{Michael Schwarz}, {and} \bibinfo{person}{Stefan Mangard}.} \bibinfo{year}{2016}\natexlab{}.
\newblock \showarticletitle{{DRAMA:} Exploiting {DRAM} Addressing for Cross-CPU Attacks}. In \bibinfo{booktitle}{\emph{25th {USENIX} Security Symposium, {USENIX} Security 16, Austin, TX, USA, August 10-12, 2016}}.
\newblock
\urldef\tempurl%
\url{https://www.usenix.org/conference/usenixsecurity16/technical-sessions/presentation/pessl}
\showURL{%
\tempurl}


\bibitem[Razavi et~al\mbox{.}(2016)]%
        {Flip_Feng_Shui}
\bibfield{author}{\bibinfo{person}{Kaveh Razavi}, \bibinfo{person}{Ben Gras}, \bibinfo{person}{Erik Bosman}, \bibinfo{person}{Bart Preneel}, \bibinfo{person}{Cristiano Giuffrida}, {and} \bibinfo{person}{Herbert Bos}.} \bibinfo{year}{2016}\natexlab{}.
\newblock \showarticletitle{Flip Feng Shui: Hammering a Needle in the Software Stack}. In \bibinfo{booktitle}{\emph{25th {USENIX} Security Symposium, {USENIX} Security 16, Austin, TX, USA, August 10-12, 2016}}.
\newblock
\urldef\tempurl%
\url{https://www.usenix.org/conference/usenixsecurity16/technical-sessions/presentation/razavi}
\showURL{%
\tempurl}


\bibitem[Ridder et~al\mbox{.}(2025)]%
        {posthammer}
\bibfield{author}{\bibinfo{person}{Finn~De Ridder}, \bibinfo{person}{Patrick Jattke}, {and} \bibinfo{person}{Kaveh Razavi}.} \bibinfo{year}{2025}\natexlab{}.
\newblock \showarticletitle{Posthammer: Pervasive Browser-based Rowhammer Attacks with Postponed Refresh Commands}. In \bibinfo{booktitle}{\emph{34th {USENIX} Security Symposium, {USENIX} Security 2025, Seattle, WA, USA, August 13-15, 2025}}.
\newblock
\urldef\tempurl%
\url{https://www.usenix.org/conference/usenixsecurity25/presentation/de-ridder}
\showURL{%
\tempurl}


\bibitem[Saileshwar et~al\mbox{.}(2022)]%
        {randomized_row_swap}
\bibfield{author}{\bibinfo{person}{Gururaj Saileshwar}, \bibinfo{person}{Bolin Wang}, \bibinfo{person}{Moinuddin~K. Qureshi}, {and} \bibinfo{person}{Prashant~J. Nair}.} \bibinfo{year}{2022}\natexlab{}.
\newblock \showarticletitle{Randomized row-swap: mitigating Row Hammer by breaking spatial correlation between aggressor and victim rows}. In \bibinfo{booktitle}{\emph{{ASPLOS} '22: 27th {ACM} International Conference on Architectural Support for Programming Languages and Operating Systems, Lausanne, Switzerland, 28 February 2022 - 4 March 2022}}.
\newblock
\urldef\tempurl%
\url{https://doi.org/10.1145/3503222.3507716}
\showURL{%
\tempurl}


\bibitem[Seaborn and Dullien(2015)]%
        {seaborn2015exploiting}
\bibfield{author}{\bibinfo{person}{Mark Seaborn} {and} \bibinfo{person}{Thomas Dullien}.} \bibinfo{year}{2015}\natexlab{}.
\newblock \showarticletitle{Exploiting the DRAM rowhammer bug to gain kernel privileges}.
\newblock \bibinfo{journal}{\emph{Black Hat}} \bibinfo{volume}{15}, \bibinfo{number}{71} (\bibinfo{year}{2015}), \bibinfo{pages}{2}.
\newblock


\bibitem[Tatar et~al\mbox{.}(2018)]%
        {Throwhammer}
\bibfield{author}{\bibinfo{person}{Andrei Tatar}, \bibinfo{person}{Radhesh~Krishnan Konoth}, \bibinfo{person}{Elias Athanasopoulos}, \bibinfo{person}{Cristiano Giuffrida}, \bibinfo{person}{Herbert Bos}, {and} \bibinfo{person}{Kaveh Razavi}.} \bibinfo{year}{2018}\natexlab{}.
\newblock \showarticletitle{Throwhammer: Rowhammer Attacks over the Network and Defenses}. In \bibinfo{booktitle}{\emph{Proceedings of the 2018 {USENIX} Annual Technical Conference, {USENIX} {ATC} 2018, Boston, MA, USA, July 11-13, 2018}}.
\newblock
\urldef\tempurl%
\url{https://www.usenix.org/conference/atc18/presentation/tatar}
\showURL{%
\tempurl}


\bibitem[Tobah et~al\mbox{.}(2022)]%
        {SpecHamme}
\bibfield{author}{\bibinfo{person}{Youssef Tobah}, \bibinfo{person}{Andrew Kwong}, \bibinfo{person}{Ingab Kang}, \bibinfo{person}{Daniel Genkin}, {and} \bibinfo{person}{Kang~G. Shin}.} \bibinfo{year}{2022}\natexlab{}.
\newblock \showarticletitle{SpecHammer: Combining Spectre and Rowhammer for New Speculative Attacks}. In \bibinfo{booktitle}{\emph{43rd {IEEE} Symposium on Security and Privacy, {SP} 2022, San Francisco, CA, USA, May 22-26, 2022}}.
\newblock
\urldef\tempurl%
\url{https://doi.org/10.1109/SP46214.2022.9833802}
\showURL{%
\tempurl}


\bibitem[Tobah et~al\mbox{.}(2024)]%
        {Gogogadget}
\bibfield{author}{\bibinfo{person}{Youssef Tobah}, \bibinfo{person}{Andrew Kwong}, \bibinfo{person}{Ingab Kang}, \bibinfo{person}{Daniel Genkin}, {and} \bibinfo{person}{Kang~G. Shin}.} \bibinfo{year}{2024}\natexlab{}.
\newblock \showarticletitle{Go Go Gadget Hammer: Flipping Nested Pointers for Arbitrary Data Leakage}. In \bibinfo{booktitle}{\emph{Proceedings of the 33rd {USENIX} Security Symposium ({USENIX} Security 2024)}}.
\newblock
\urldef\tempurl%
\url{https://www.usenix.org/conference/usenixsecurity24/presentation/tobah}
\showURL{%
\tempurl}


\bibitem[van~der Veen et~al\mbox{.}(2016)]%
        {Drammer}
\bibfield{author}{\bibinfo{person}{Victor van~der Veen}, \bibinfo{person}{Yanick Fratantonio}, \bibinfo{person}{Martina Lindorfer}, \bibinfo{person}{Daniel Gruss}, \bibinfo{person}{Cl{\'{e}}mentine Maurice}, \bibinfo{person}{Giovanni Vigna}, \bibinfo{person}{Herbert Bos}, \bibinfo{person}{Kaveh Razavi}, {and} \bibinfo{person}{Cristiano Giuffrida}.} \bibinfo{year}{2016}\natexlab{}.
\newblock \showarticletitle{Drammer: Deterministic Rowhammer Attacks on Mobile Platforms}. In \bibinfo{booktitle}{\emph{Proceedings of the 2016 {ACM} {SIGSAC} Conference on Computer and Communications Security, Vienna, Austria, October 24-28, 2016}}.
\newblock
\urldef\tempurl%
\url{https://doi.org/10.1145/2976749.2978406}
\showURL{%
\tempurl}


\bibitem[Wang et~al\mbox{.}(2020)]%
        {dramdig}
\bibfield{author}{\bibinfo{person}{Minghua Wang}, \bibinfo{person}{Zhi Zhang}, \bibinfo{person}{Yueqiang Cheng}, {and} \bibinfo{person}{Surya Nepal}.} \bibinfo{year}{2020}\natexlab{}.
\newblock \showarticletitle{DRAMDig: {A} Knowledge-assisted Tool to Uncover {DRAM} Address Mapping}. In \bibinfo{booktitle}{\emph{57th {ACM/IEEE} Design Automation Conference, {DAC} 2020, San Francisco, CA, USA, July 20-24, 2020}}.
\newblock
\urldef\tempurl%
\url{https://doi.org/10.1109/DAC18072.2020.9218599}
\showURL{%
\tempurl}


\bibitem[Wi et~al\mbox{.}(2023)]%
        {shadow}
\bibfield{author}{\bibinfo{person}{Minbok Wi}, \bibinfo{person}{Jaehyun Park}, \bibinfo{person}{Seoyoung Ko}, \bibinfo{person}{Michael~Jaemin Kim}, \bibinfo{person}{Nam~Sung Kim}, \bibinfo{person}{Eojin Lee}, {and} \bibinfo{person}{Jung~Ho Ahn}.} \bibinfo{year}{2023}\natexlab{}.
\newblock \showarticletitle{{SHADOW:} Preventing Row Hammer in {DRAM} with Intra-Subarray Row Shuffling}. In \bibinfo{booktitle}{\emph{{IEEE} International Symposium on High-Performance Computer Architecture, {HPCA} 2023, Montreal, QC, Canada, February 25 - March 1, 2023}}.
\newblock
\urldef\tempurl%
\url{https://doi.org/10.1109/HPCA56546.2023.10070966}
\showURL{%
\tempurl}


\bibitem[Woo et~al\mbox{.}(2023)]%
        {scalable_row_swap}
\bibfield{author}{\bibinfo{person}{Jeonghyun Woo}, \bibinfo{person}{Gururaj Saileshwar}, {and} \bibinfo{person}{Prashant~J. Nair}.} \bibinfo{year}{2023}\natexlab{}.
\newblock \showarticletitle{Scalable and Secure Row-Swap: Efficient and Safe Row Hammer Mitigation in Memory Systems}. In \bibinfo{booktitle}{\emph{{IEEE} International Symposium on High-Performance Computer Architecture, {HPCA} 2023, Montreal, QC, Canada, February 25 - March 1, 2023}}.
\newblock
\urldef\tempurl%
\url{https://doi.org/10.1109/HPCA56546.2023.10070999}
\showURL{%
\tempurl}


\bibitem[Xiao et~al\mbox{.}(2016)]%
        {one_Bit_Flips_One_Cloud_Flops}
\bibfield{author}{\bibinfo{person}{Yuan Xiao}, \bibinfo{person}{Xiaokuan Zhang}, \bibinfo{person}{Yinqian Zhang}, {and} \bibinfo{person}{Radu Teodorescu}.} \bibinfo{year}{2016}\natexlab{}.
\newblock \showarticletitle{One Bit Flips, One Cloud Flops: Cross-VM Row Hammer Attacks and Privilege Escalation}. In \bibinfo{booktitle}{\emph{25th {USENIX} Security Symposium, {USENIX} Security 16, Austin, TX, USA, August 10-12, 2016}}.
\newblock
\urldef\tempurl%
\url{https://www.usenix.org/conference/usenixsecurity16/technical-sessions/presentation/xiao}
\showURL{%
\tempurl}


\bibitem[Yao et~al\mbox{.}(2020)]%
        {deephammer}
\bibfield{author}{\bibinfo{person}{Fan Yao}, \bibinfo{person}{Adnan~Siraj Rakin}, {and} \bibinfo{person}{Deliang Fan}.} \bibinfo{year}{2020}\natexlab{}.
\newblock \showarticletitle{DeepHammer: Depleting the Intelligence of Deep Neural Networks through Targeted Chain of Bit Flips}. In \bibinfo{booktitle}{\emph{29th {USENIX} Security Symposium, {USENIX} Security 2020, August 12-14, 2020}}.
\newblock
\urldef\tempurl%
\url{https://www.usenix.org/conference/usenixsecurity20/presentation/yao}
\showURL{%
\tempurl}


\bibitem[Zhang et~al\mbox{.}(2025)]%
        {erebor}
\bibfield{author}{\bibinfo{person}{Chuqi Zhang}, \bibinfo{person}{Rahul Priolkar}, \bibinfo{person}{Yuancheng Jiang}, \bibinfo{person}{Yuan Xiao}, \bibinfo{person}{Mona Vij}, \bibinfo{person}{Zhenkai Liang}, {and} \bibinfo{person}{Adil Ahmad}.} \bibinfo{year}{2025}\natexlab{}.
\newblock \showarticletitle{Erebor: {A} Drop-In Sandbox Solution for Private Data Processing in Untrusted Confidential Virtual Machines}. In \bibinfo{booktitle}{\emph{Proceedings of the Twentieth European Conference on Computer Systems, EuroSys 2025, Rotterdam, The Netherlands, 30 March 2025 - 3 April 2025}}.
\newblock
\urldef\tempurl%
\url{https://doi.org/10.1145/3689031.3717464}
\showURL{%
\tempurl}


\bibitem[Zhang et~al\mbox{.}(2024b)]%
        {hitchhiker}
\bibfield{author}{\bibinfo{person}{Chuqi Zhang}, \bibinfo{person}{Jun Zeng}, \bibinfo{person}{Yiming Zhang}, \bibinfo{person}{Adil Ahmad}, \bibinfo{person}{Fengwei Zhang}, \bibinfo{person}{Hai Jin}, {and} \bibinfo{person}{Zhenkai Liang}.} \bibinfo{year}{2024}\natexlab{b}.
\newblock \showarticletitle{The HitchHiker's Guide to High-Assurance System Observability Protection with Efficient Permission Switches}. In \bibinfo{booktitle}{\emph{Proceedings of the 2024 on {ACM} {SIGSAC} Conference on Computer and Communications Security, {CCS} 2024, Salt Lake City, UT, USA, October 14-18, 2024}}.
\newblock
\urldef\tempurl%
\url{https://doi.org/10.1145/3658644.3690188}
\showURL{%
\tempurl}


\bibitem[Zhang et~al\mbox{.}(2024a)]%
        {Sok}
\bibfield{author}{\bibinfo{person}{Zhi Zhang}, \bibinfo{person}{Decheng Chen}, \bibinfo{person}{Jiahao Qi}, \bibinfo{person}{Yueqiang Cheng}, \bibinfo{person}{Shijie Jiang}, \bibinfo{person}{Yiyang Lin}, \bibinfo{person}{Yansong Gao}, \bibinfo{person}{Surya Nepal}, \bibinfo{person}{Yi Zou}, \bibinfo{person}{Jiliang Zhang}, {and} \bibinfo{person}{Yang Xiang}.} \bibinfo{year}{2024}\natexlab{a}.
\newblock \showarticletitle{SoK: Rowhammer on Commodity Operating Systems}. In \bibinfo{booktitle}{\emph{Proceedings of the 19th {ACM} Asia Conference on Computer and Communications Security, {ASIA} {CCS} 2024, Singapore, July 1-5, 2024}}.
\newblock
\urldef\tempurl%
\url{https://doi.org/10.1145/3634737.3656998}
\showURL{%
\tempurl}


\bibitem[Zhang et~al\mbox{.}(2020a)]%
        {pthammer}
\bibfield{author}{\bibinfo{person}{Zhi Zhang}, \bibinfo{person}{Yueqiang Cheng}, \bibinfo{person}{Dongxi Liu}, \bibinfo{person}{Surya Nepal}, \bibinfo{person}{Zhi Wang}, {and} \bibinfo{person}{Yuval Yarom}.} \bibinfo{year}{2020}\natexlab{a}.
\newblock \showarticletitle{PThammer: Cross-User-Kernel-Boundary Rowhammer through Implicit Accesses}. In \bibinfo{booktitle}{\emph{53rd Annual {IEEE/ACM} International Symposium on Microarchitecture, {MICRO} 2020, Athens, Greece, October 17-21, 2020}}.
\newblock
\urldef\tempurl%
\url{https://doi.org/10.1109/MICRO50266.2020.00016}
\showURL{%
\tempurl}


\bibitem[Zhang et~al\mbox{.}(2020b)]%
        {ghostknight}
\bibfield{author}{\bibinfo{person}{Zhi Zhang}, \bibinfo{person}{Yueqiang Cheng}, \bibinfo{person}{Yinqian Zhang}, {and} \bibinfo{person}{Surya Nepal}.} \bibinfo{year}{2020}\natexlab{b}.
\newblock \showarticletitle{GhostKnight: Breaching Data Integrity via Speculative Execution}.
\newblock \bibinfo{journal}{\emph{CoRR}}  \bibinfo{volume}{abs/2002.00524} (\bibinfo{year}{2020}).
\newblock
\urldef\tempurl%
\url{https://arxiv.org/abs/2002.00524}
\showURL{%
\tempurl}


\bibitem[Zhang et~al\mbox{.}(2023)]%
        {implicit_hammer}
\bibfield{author}{\bibinfo{person}{Zhi Zhang}, \bibinfo{person}{Wei He}, \bibinfo{person}{Yueqiang Cheng}, \bibinfo{person}{Wenhao Wang}, \bibinfo{person}{Yansong Gao}, \bibinfo{person}{Dongxi Liu}, \bibinfo{person}{Kang Li}, \bibinfo{person}{Surya Nepal}, \bibinfo{person}{Anmin Fu}, {and} \bibinfo{person}{Yi Zou}.} \bibinfo{year}{2023}\natexlab{}.
\newblock \showarticletitle{Implicit Hammer: Cross-Privilege-Boundary Rowhammer Through Implicit Accesses}.
\newblock \bibinfo{journal}{\emph{{IEEE} Trans. Dependable Secur. Comput.}} \bibinfo{volume}{20}, \bibinfo{number}{5} (\bibinfo{year}{2023}), \bibinfo{pages}{3716--3733}.
\newblock
\urldef\tempurl%
\url{https://doi.org/10.1109/TDSC.2022.3214666}
\showDOI{\tempurl}


\bibitem[Zhang et~al\mbox{.}(2021)]%
        {bitmine}
\bibfield{author}{\bibinfo{person}{Zhi Zhang}, \bibinfo{person}{Wei He}, \bibinfo{person}{Yueqiang Cheng}, \bibinfo{person}{Wenhao Wang}, \bibinfo{person}{Yansong Gao}, \bibinfo{person}{Minghua Wang}, \bibinfo{person}{Kang Li}, \bibinfo{person}{Surya Nepal}, {and} \bibinfo{person}{Yang Xiang}.} \bibinfo{year}{2021}\natexlab{}.
\newblock \showarticletitle{BitMine: An End-to-End Tool for Detecting Rowhammer Vulnerability}.
\newblock \bibinfo{journal}{\emph{{IEEE} Trans. Inf. Forensics Secur.}}  \bibinfo{volume}{16} (\bibinfo{year}{2021}), \bibinfo{pages}{5167--5181}.
\newblock
\urldef\tempurl%
\url{https://doi.org/10.1109/TIFS.2021.3124728}
\showDOI{\tempurl}


\bibitem[Zhang et~al\mbox{.}(2020c)]%
        {Leveraging_EM_Side_Channel_Information_to_Detect_Rowhammer_Attacks}
\bibfield{author}{\bibinfo{person}{Zhenkai Zhang}, \bibinfo{person}{Zihao Zhan}, \bibinfo{person}{Daniel Balasubramanian}, \bibinfo{person}{Bo Li}, \bibinfo{person}{Peter Volgyesi}, {and} \bibinfo{person}{Xenofon Koutsoukos}.} \bibinfo{year}{2020}\natexlab{c}.
\newblock \showarticletitle{Leveraging {EM} Side-Channel Information to Detect Rowhammer Attacks}. In \bibinfo{booktitle}{\emph{2020 IEEE Symposium on Security and Privacy (SP)}}.
\newblock
\urldef\tempurl%
\url{https://doi.org/10.1109/SP40000.2020.00060}
\showURL{%
\tempurl}


\end{thebibliography}

\end{document}